\documentclass[longauth]{aa}  
\usepackage{ulem}
\usepackage{graphicx}
\usepackage{txfonts}
\usepackage{xcolor}
\usepackage[colorlinks=true,linkcolor=red,anchorcolor=blue,citecolor=blue,filecolor=black,menucolor=black,runcolor=black,urlcolor=blue]{hyperref}

\newcommand{\Msyr}{\,$M_\odot$\,yr$^{-1}$\xspace}
\newcommand{\Ms}{\,$M_\odot$\xspace}
\newcommand{\arcs}{$\arcsec$\xspace}
\newcommand{\micron}{\,$\mu$m\xspace}
\newcommand{\Myr}{\,Myr\xspace}

\newcommand{\uJy}{\,$\mu$Jy\xspace}

\newcommand{\simi}{$\sim$\,}
\newcommand{\pmm}{\,$\pm$\,}
\newcommand{\EXP}[1]{\,$\times$\,10$^{#1}$}
\newcommand{\Mstar}{$M_\mathrm{*}$\xspace}

\begin{document}


    \title{JWST/MIRI unveils the stellar component of the GN20 dusty galaxy overdensity at $z$\,=\,4.05}
   

   \author{A. Crespo G\'omez\inst{\ref{inst:CAB}} \and L. Colina\inst{\ref{inst:CAB}} 
   \and J. \'Alvarez-M\'arquez\inst{\ref{inst:CAB}} \and A. Bik\inst{\ref{inst:Stockholm}}
   \and L. Boogaard\inst{\ref{inst:MPIA}} 
   \and G. {\"O}stlin\inst{\ref{inst:Stockholm}}  \and  F. Pei{\ss}ker\inst{\ref{inst:Köln}}
   \and F. Walter\inst{\ref{inst:MPIA}} 
   \and A. Labiano\inst{\ref{inst:CAB-ESAC},\ref{inst:Telespazio}} 
   \and P. G. P\'erez-Gonz\'alez\inst{\ref{inst:CAB}} 
   \and T. R. Greve\inst{\ref{inst:DTU}, \ref{inst:DAWN}, \ref{inst:UCL}} 
   \and G. Wright\inst{\ref{inst:UKATC}}
   \and A. Alonso-Herrero\inst{\ref{inst:CAB-ESAC}}
   \and K. I. Caputi\inst{\ref{inst:Groningen}}
   \and L. Costantin \inst{\ref{inst:CAB}} 
   \and A. Eckart\inst{\ref{inst:Köln}} 
   \and M. Garc\'ia-Mar\'in\inst{\ref{inst:ESA}}
   \and S. Gillman\inst{\ref{inst:DTU}, \ref{inst:DAWN}}
   \and J. Hjorth\inst{\ref{inst:DARK}} 
   \and E. Iani\inst{\ref{inst:Groningen}} 
   \and D. Langeroodi\inst{\ref{inst:DARK}} 
   \and J. P. Pye\inst{\ref{inst:Leicester}}
   \and P. Rinaldi\inst{\ref{inst:Groningen}}
   \and T. Tikkanen\inst{\ref{inst:Leicester}} 
   \and P. van der Werf\inst{\ref{inst:Leiden}} 
    \and P. O. Lagage\inst{\ref{inst:AIM}} 
    \and E. F. van Dishoeck\inst{\ref{inst:Leiden}}  
   }

   \institute{Centro de Astrobiolog\'{\i}a (CAB), CSIC-INTA, Ctra. de Ajalvir km 4, Torrej\'on de Ardoz, E-28850, Madrid, Spain\\  \email{acrespo@cab.inta-csic.es} \label{inst:CAB}
   \and Department of Astronomy, Stockholm University, Oscar Klein Centre, AlbaNova University Centre, 106 91 Stockholm, Sweden \label{inst:Stockholm}
    \and Max-Planck-Institut f\"ur Astronomie, K\"onigstuhl 17, 69117 Heidelberg, Germany\label{inst:MPIA}
    \and I.Physikalisches Institut der Universit\"at zu K\"oln, Z\"ulpicher Str. 77, 50937 K\"oln, Germany \label{inst:Köln}
    \and Centro de Astrobiolog\'ia (CAB), CSIC-INTA, Camino Viejo del Castillo s/n, 28692 Villanueva de la Ca\~{n}ada, Madrid, Spain \label{inst:CAB-ESAC}
    \and Telespazio UK for the European Space Agency, ESAC, Camino Bajo del Castillo s/n, 28692 Villanueva de la Ca\~{n}ada, Spain \label{inst:Telespazio}
    \and Cosmic Dawn Center, DTU Space, Technical University of Denmark, Elektrovej 327, 2800 Kgs. Lyngby, Denmark \label{inst:DTU}
    \and Cosmic Dawn Centre, Copenhagen, Denmark \label{inst:DAWN} 
    \and Department of Physics and Astronomy, University College London, Gower Place, London WC1E 6BT, UK \label{inst:UCL}
    \and UK Astronomy Technology Centre, Royal Observatory Edinburgh, Blackford Hill, Edinburgh EH9 3HJ, UK \label{inst:UKATC} 
    \and Kapteyn Astronomical Institute, University of Groningen, P.O. Box 800, 9700 AV Groningen, The Netherlands \label{inst:Groningen} 
    \and European Space Agency, Space Telescope Science Institute, Baltimore, Maryland, USA \label{inst:ESA} 
    \and DARK, Niels Bohr Institute, University of Copenhagen, Jagtvej 128, 2200 Copenhagen, Denmark \label{inst:DARK}
    \and School of Physics \& Astronomy, Space Park Leicester, University of Leicester, 92 Corporation Road, Leicester LE4 5SP, UK \label{inst:Leicester} 
    \and Leiden Observatory, Leiden University, PO Box 9513, 2300 RA Leiden, The Netherlands \label{inst:Leiden}
    %
    \and Universit\'e Paris-Saclay, Universit\'e Paris Cit\'e, CEA, CNRS, AIM, F-91191 Gif-sur-Yvette, France \label{inst:AIM}
    }

   \date{Received ; accepted}


 

\abstract{Dusty star-forming galaxies (DSFGs) at $z$\,>\,2 have been commonly observed in overdense regions, where the merging processes and large halo masses induce rapid gas accretion, triggering star formation rates (SFRs) up to \simi1000\Msyr. Despite the importance of these DSFGs for understanding  star formation in the early Universe, their stellar distributions, traced by the near-infrared (near-IR) emission, had remained spatially unresolved until the arrival of the JWST. In this work, we present, for the first time, a spatially resolved morphological analysis of the rest-frame near-IR (\simi1.1\,$-$\,3.5\micron) emission in DSFGs traced with the JWST/MIRI F560W, F770W, F1280W, and F1800W filters. In particular, we studied the mature stellar component for the three DSFGs and a Lyman-break galaxy (LBG) present in an overdensity at $z$\,=\,4.05. Moreover, we used these rest-frame near-IR images along with ultraviolet (UV) and (sub)-mm ancillary photometric data to model their spectral energy distributions (SEDs) and extract their main physical properties (e.g. \,\Mstar, SFR, $A_\mathrm{V}$). 
The sub-arcsec resolution images from the JWST have revealed that the light distributions in these galaxies present a wide range of morphologies, from disc-like to compact and clump-dominated structures. Two DSFGs and the LBG are classified as late-type galaxies (LTGs) according to non-parametric morphological indices, while the remaining DSFG is an early-type galaxy (ETG). These near-IR structures contrast with their ultraviolet emission, which is diffuse and, in GN20 and GN20.2b, off-centred by \simi4\,kpc. This result suggests that star formation takes place across the entire galaxy, while the UV light traces only those regions where the otherwise high internal extinction decreases significantly. 
The SED fitting analysis yields large SFRs (\simi300\,$-$2500\Msyr), large stellar masses (\Mstar{}=\,(0.24\,$-$\,1.79)\EXP{11}\Ms), and high integrated extinction values ($A_\mathrm{V}$\,=\,0.8$-$1.5\,mag) for our  galaxies. In particular, we observe that GN20 dominates the total SFR with a value 2550\pmm150\Msyr, while GN20.2b has the highest stellar mass (\Mstar{}=\,(2.2\pmm1.4)\EXP{11}\Ms). The two DSFGs classified as LTGs (GN20 and GN20.2a) have a high specific SFR (sSFR\,>\,30\,Gyr$^{-1}$), placing them above the star-forming main sequence (SFMS) at z\,\simi4 by \simi0.5\,dex; whereas the ETG (i.e. \,GN20.2b) is compatible with the high-mass end of the main sequence. In comparison with other DSFGs in overdensities at $z$\,\simi2\,$-$\,7, we observe that our objects present similar SFRs, depletion times, and projected separations. Nevertheless, the sizes computed for GN20 and GN20.2a are up to two times larger than those of isolated galaxies observed in CEERS and ALMA-HUDF at similar redshifts. We interpret this difference in size as an effect of rapid growth induced by the dense environment.}

\titlerunning{--}
\maketitle

\section{Introduction}
\label{sec:Introduction}

Dusty star-forming galaxies (DSFGs) are massive and extremely infrared-bright objects that are characterised by their intense starburst episodes, with star formation rates (SFRs) greater than 200\Msyr, and high level of dust content that obscures most of their ultraviolet (UV) and optical emission (see \citealt{Casey+14}, for a review). Many of these DSFGs have  also been classified as sub-millimetre galaxies (SMGs) due to their strong emission at these wavelengths \citep{Smail+97,Barger+98,Hughes+98, Borys+03}. These starburst galaxies are key to understanding star formation history (SFH) across the Universe \citep[see][for a review]{Madau&Dickinson+14} as they dominate the cosmic star formation at $z$~\simi4 \citep{Perez-Gonzalez+05,Zavala+21}. Additionally, DSFGs are also important to understand the galaxy evolution since they are thought to be the progenitors of massive quiescent galaxies at z\,\simi2 \citep{Valentino+20}. Although the DSFGs have a redshift distribution that peaks at $z$\,\simi2\,$-$\,3 \citep{Chapman+05}, a significant number of these objects have been found at $z$\,$>$\,4 \citep{Walter+12,Riechers+13,Strandet+17,Zavala+21,Zavala+23}.

Different scenarios are proposed to explain the characteristic extreme SFRs of these objects. Many DSFGs are thought to be dominated by secular evolution, where high SFR is caused by the infall of cold molecular gas into gravitationally unstable discs, which creates star-forming clumps that eventually migrate to the centres of these galaxies \citep{Dekel+09,Ceverino+10,Wuyts+12,Adamo+13,Mandelker+14,Krumholz+18}. The other scenario explains their large SFR as a consequence of gravitational interactions (i.e. mergers) which fuels these galaxies with molecular gas and triggers their star formation \citep{Cochrane+21,Spilker+22,Alvarez-Marquez+23,Fuentealba-Fuentes+24}. At $z$\,$>$\,2, DSFGs have been discovered to be often part of overdensities and proto-clusters \citep{Daddi+09,Riechers+10,Oteo+18,Pavesi+18,Drake+20}, which favour the gravitational interactions. This connection between DSFGs and galaxy overdensities has been shown to be more common in the early Universe \citep{Smolcic+17,Lewis+18,Hashimoto+23,Arribas+24}. Studying the physical mechanisms driving the DSFGs in overdensities in detail  will provide valuable information on the halo properties and processes (e.g. \,gas-cooling, dark matter mass, gas accretion, and gravitational interactions), which is key to understanding the galaxy formation and evolution at the early stages of the Universe.

To date, most of the spatially resolved analyses carried out in DSFGs at $z$\,$>$\,4 have been focused on their UV and (sub)-mm emission, tracing their very young stellar population, as well as the molecular gas and dust distributions \citep{Carilli+10,Hodge+15,Pavesi+18,Gomez-Guijarro+19,Hodge+19,Tadaki+20}. However, the available IR instruments did not allow to spatially resolve these objects, leaving the structure of the host galaxy, traced by mature stellar populations, unknown. The arrival of \textit{James Webb Space Telescope} (JWST, \citealt{Gardner+23}) has altered this scenario by offering the opportunity to observe the rest-frame optical and near-IR emission of these galaxies with unprecedented sensitivity and angular resolution \citep{Alvarez-Marquez+23,Colina+23,Zavala+23}. In particular, the rest-frame near-IR light captured by the JWST at $z$\,\simi4$-$5 is emitted by red giants and supergiant stars created $>$100\,Myr ago, tracing the mature stellar population of high-$z$ galaxies. In addition, IR wavelengths are less affected by dust extinction than the optical and UV, which makes the mid-IR filters of MIRI \citep{Rieke+15,Wright+15,Wright+23}  the optimal option for investigating the true stellar structure in the highly obscured DSFGs.

In this work, we make use of the great capabilities of the JWST MIRI instrument to analyse the stellar component of four spectroscopically-confirmed galaxies (with three of them identified as DSFGs) in an overdensity at $z$\,\simi4 \citep{Daddi+09}. Two of these DSFGs (GN20 and GN20.2a) were previously discovered by \citet{Pope+05} and classified as SMGs based on\ their extreme SCUBA 850\micron fluxes. Later, \citet{Daddi+09} confirmed their redshift at $z$\,=\,4.05 and found another tentative object (i.e. \,GN20.2b) at the same redshift, as a serendipitous detection of CO(4-3) emission with the Plateau de Bure Interferometer (PdBI). The authors also discovered a Lyman-break galaxy (LBG), named BD29079, at this redshift with Keck spectroscopic data. All these galaxies are within \simi20\arcs of each other, which at $z$\,=\,4.05 corresponds to a projected separation of \simi140\,kpc. The IR-to-radio spectral energy distribution (SED) analysis of these SMGs revealed very high IR luminosities ($>$\,1\EXP{13}\,$L_\odot$), whereas their CO kinematics yields dynamical masses larger than 10$^{10}$\Ms. Using PdBI continuum data along with ancillary multiwavelength (UV-to-radio) photometry, \citet{Tan+13} and \citet{Tan+14} derived that the SMGs in the overdensity are quite massive ($\log$(\Mstar/\Ms)\,\simi10\,$-$\,11) and present large SFRs (\simi500\,$-$\,2000\Msyr), classifying them as starburst galaxies. The high resolution of the CO(2-1) emission maps obtained from the VLA allowed \citet{Hodge+13} to find significant offsets (\simi0.5\,$-$\,1\arcs) between the CO emission and their rest-frame UV counterparts, which suggest either the existence of projected near galaxies emitting this UV light or the presence of a large dust attenuation in the nuclear region of these galaxies. Aside from these four spectroscopically confirmed objects at this redshift, \citet{Daddi+09} found another ten B-dropout galaxies with photometric redshifts $z$\,\simi4 within 25\arcs of GN20. None of these objects show CO(2-1) emission \citep{Hodge+13} and, to date, they have not been spectroscopically confirmed. Therefore, in this work, we only focus on the 3 DSFGs and the LBG with spectroscopic redshift $z$\,=\,4.05.

The JWST/MIRI filters used in this work, from F560W to F1800W, trace the rest near-IR emission (\simi1.1\,$-$\,3.5\micron), filling the gap between the rest-frame UV and the (sub)-mm data and allowing us to observe the stellar structure of these galaxies at sub-arcsec resolution (\simi0.25\arcs). The paper is organised as follows. Section~\ref{sec:2.General_observations} presents the JWST and HST data used in this work. The morphological analysis based on the rest-frame near-IR emission and the SED fitting carried out in this work is introduced in Sect.~\ref{sec:3.Analysis}. In Section~\ref{sec:4.Results}, we characterise the morphological structure of our sample and present the physical properties derived from the SED fitting. These results are then compared with isolated galaxies at $z$\,\simi4 and with a sample of SMGs at $z$\,$<$\,5. The main conclusions and results are summarised in Section~\ref{sec:5.Summary}.

Throughout the paper we assume a Chabrier initial mass function (IMF, \citealt{Chabrier+03}) and a flat $\Lambda$CDM cosmology, with $\Omega_\mathrm{m}$\,=\,0.31 and H$_0$\,=\,67.7\,km\,s$^{-1}$\,Mpc$^{-1}$ \citep{PlanckCollaboration18VI}. For this cosmology, 1 arcsec corresponds to 7.05\,kpc at $z$\,=\,4.05 and the luminosity distance is $D_\mathrm{L}$\,=\,37.13\,Gpc. 

\section{Observation and data processing}
\label{sec:2.General_observations}

\subsection{JWST/MIRIM data and calibration}
\label{subsec:JWST_data}

The JWST images of the GN20 field were obtained on November 23-24, 2022 using the  MIRI imager \citep[MIRIM,][]{Bouchet+15} in the F560W, F770W, F1280W, and F1800W filters  as part of the MIRI European Consortium Guaranteed Time (program ID 1264, PI: Colina, L.). The observations were taken using the FASTR1 read-out mode for total integration time of 1498 seconds in a five-dither medium-size cycling pattern for the F560W, F770W, and F1280W filters. We obtained the F1800W image as a simultaneous MIRI observation during the acquisition of background-dedicated MRS data (see \citealt{Bik+24} for details). Therefore, it presents different observational setup than the rest of the MIRI data with an integration time of 1911 seconds and the two-point dither pattern with the FASTR1 read-out mode.


All these MIRIM images have been calibrated using the JWST pipeline (v1.12.0) with the context 1140 of the Calibration Reference Data System (CRDS). This CRDS includes the most recent photometric calibrations considering the temporal evolution and aperture corrections taking into account the point-spread function (PSF) cruciform. In addition to the general procedure, additional steps have been applied to correct for striping and background gradients (see details in \citealt{Alvarez-Marquez+23}). After dithering, the final MIRI images have a scale of 0.06$\arcsec$ per pixel. The spatial resolutions of these images are, according to the JWST documentation\footnote{https://jwst-docs.stsci.edu/jwst-mid-infrared-instrument/miri-performance/miri-point-spread-functions}, FWHM\,=\,0.207\arcs, 0.269\arcs, 0.420\arcs and 0.591\arcs for the F560W, F770W, F1280W, and F1800W filters, respectively.

\subsection{HST ancillary data}
\label{subsec:ancillary_data}

In addition to the MIRI data, we retrieved ancillary HST images to trace the rest-frame UV counterparts of the objects in our sample. These images were downloaded from the \textit{Rainbow} Cosmological Surveys Database through its web interface \textit{Rainbow Navigator} \citep{Barro+11a}. Specifically, we downloaded the ACS/WFC F435W, F606W, F770W, F814W, and F850LP images, along with the WFC3$\_$IR F105W, F140W, F125W, and F160W images from the CANDELS/SHARDS catalogue presented in \citet{Barro+19}. The spatial resolution of these images varies between FWHM\,\simi0.1$-$0.2\arcs, whereas their 5$\sigma$ depth is \simi27\,mag (see Table\,2 from \citealt{Barro+19} for further details).

\subsection{JWST-HST astrometry}
\label{subsec:astrometry}

Before analysing the data, we set the HST and MIRI images in the same coordinate system. We used the available \textit{Gaia} DR3 \citep{GaiaCollaboration+23} stars within the FoV (usually three or four of them) to align the images, yielding a typical uncertainty that is smaller than 40\,mas in the final absolute positioning. The typical angular resolutions of the far infrared (FIR) and sub-millimetre (sub-mm) photometric data (e.g.  IRAC, Herschel, VLA) used in the SED fitting (see Sect.~\ref{subsec:sed_fitting}) have an FWHM\,$>$\,1.5\arcs \citep{Liu+19}, more than 25 times larger than the uncertainties derived from the astrometry alignment. Therefore, we can safely assume that all these photometric values correspond to our galaxies.

\section{Analysis}
\label{sec:3.Analysis}

\subsection{Surface brightness modelling and non-parametric morphology}
\label{subsec:Morph_analysis}

\begin{figure*}[!ht]
\centering
   \includegraphics[width=0.99\linewidth]{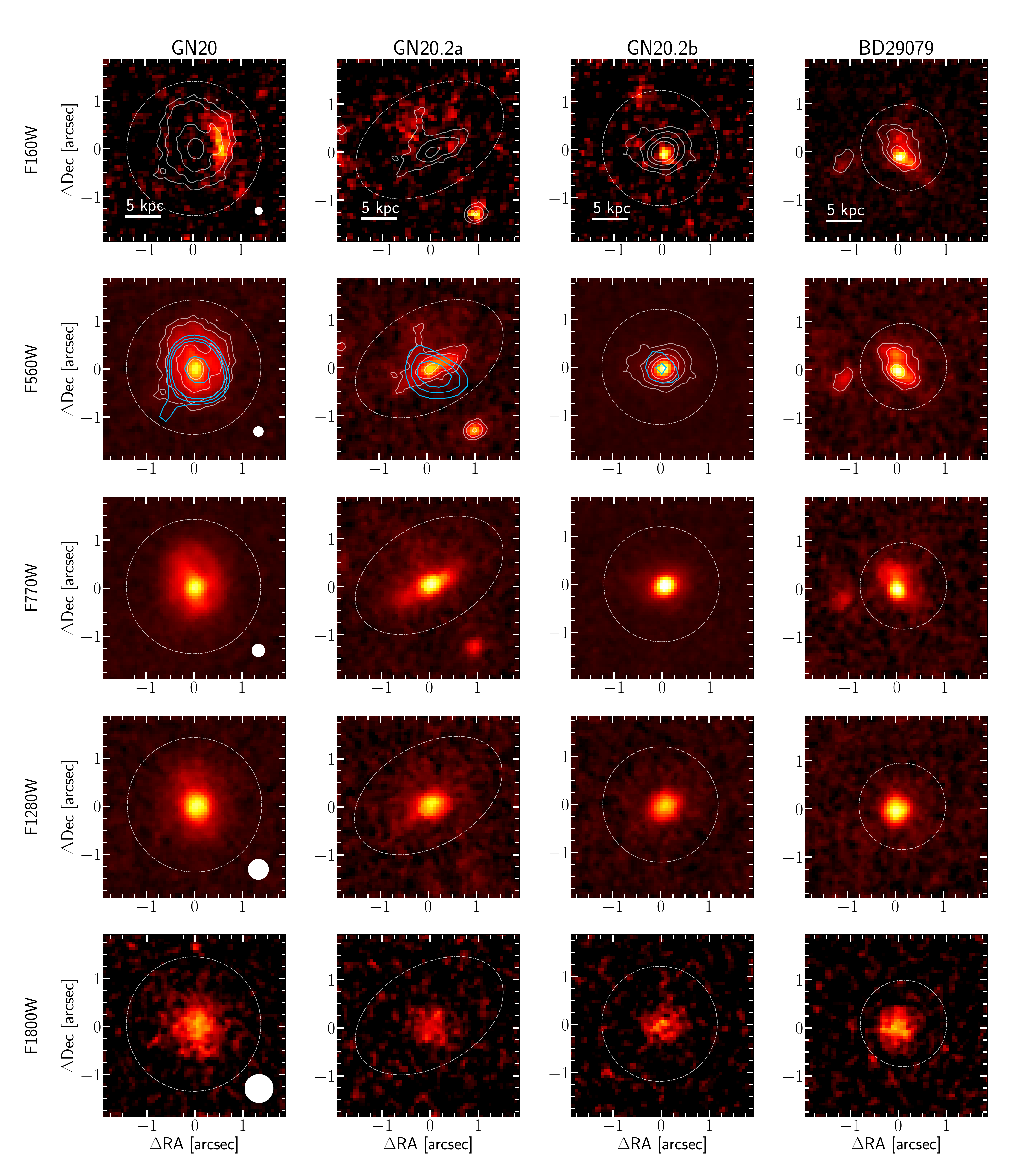}
      \caption{Rest-frame UV and near-IR for the galaxies in the sample. From top to bottom: HST/F160W, MIRI/F560W, MIRI/F770W, MIRI/F1280W, and MIRI/F1800W images for the three DSFGs and the LBG at $z$\,=\,4.05, tracing the rest-frame \simi0.3\micron, 1.1\micron, 1.5\micron, 2.5\micron, and 3.5\micron emission, respectively. White contours represent the F560W isophotes at 5$\sigma$, 10$\sigma$, 20$\sigma,$ and 35$\sigma$ levels, where $\sigma$ is the standard deviation of the background emission of each image. Dashed-line circles and ellipses represent the apertures used to extract the integrated flux for each object. Filled white circles at the lower right corner of the left-most panels show the PSF size for each image. The (0,0) positions mark the brightest pixel in the F560W image for each object. Blue contours in the second-row panels display the PdBI CO(6-5) emission for the DSFGs presented in \citet{Carilli+10} and \citet{Hodge+13}.}
         \label{fig:UV-MIRI_images}
\end{figure*}

To analyse the morphological structure of the DSFGs and the LBG, we performed parametric fits on the F560W and F770W brightness distributions, as they have the highest spatial resolution and signal-to-noise ratio (S/N, see Fig.~\ref{fig:UV-MIRI_images}). For this modelling process, we used the \texttt{Lenstronomy} code \citep{Birrer+18}, which allows us to perform an MCMC analysis to compute the associated uncertainties of the derived parameters. During the fitting procedure, we considered the empirical PSFs presented in \citet{Libralato+24} as they are a good representation of the field stars and at a better S/N. 

The different morphologies shown by our galaxies, ranging from disc- to clump-dominated objects (see Fig.~\ref{fig:UV-MIRI_images}), demand us to consider different procedures to fit their surface brightness distributions. For GN20 we followed the approach presented in \citet{Colina+23}: we considered an extended emission, modelled with a S\'ersic profile, and a point-source nucleus. As GN20.2a presents a very similar brightness distribution, we assumed the same approach for this galaxy. GN20.2b, on the other hand, shows a very compact structure. Therefore, we assumed a single S\'ersic profile to describe its brightness distribution. Table~\ref{tab:Lenstronomy} summarises the main physical parameters derived from these models. A visual inspection of the F560W image of BD29079 shows how this object is composed of three bright clumps. Therefore, we considered three point-like sources to model the near-IR emission of this galaxy. In Sect.~\ref{subsec:morph-results}, we discuss whether these clumps are better described as point-like sources or can be fitted with S\'ersic profiles.

Along with the brightness distribution modelling, we have also applied the Lucy-Richardson (LR) deconvolution algorithm \citep{Lucy+74} to the F560W images using its empirical PSF and 10\,000 iterations. The choice of such large number of iterations allow us to rule out the presence of spurious sources constructed solely on sporadic noise fluctuations (see App.~\ref{ref:high_pass_appendix} for further details). The basic idea of this approach is to enhance structures that are blended by the PSF wings of brighter sources in the related FoV. After the iterative process, we convolved the resulting delta-maps with a three-pixel Gaussian kernel filter, as described in detail in \citet{Peissker+22}. With the dampened impact of the PSF wings, these convolved maps allow us to observe the intrinsic morphology of the galaxies, revealing sub-structures that are smaller than the MIRI spatial resolution.

In addition, we also measured non-parametric indices for the near-IR images using the python package \texttt{Statmorph} \citep{Rodriguez-Gomez+19}. We used this code to obtain widely-used statistical values such as the concentration ($C$), asymmetry ($A$), Gini coefficient ($G$), and the $M_\mathrm{20}$. These statistical indices have been historically used to classify galaxies by their morphology and to determine whether they are under gravitational interaction or not \citep{Bershady+00,Lotz+04,Conselice+14}. To quantify the uncertainties associated with these indices, we performed a Monte Carlo simulation with 500 iterations, adding random noise to the original MIRI images with $\sigma$ equal to the root mean square (rms) of their background. During the procedure we masked the additional objects and diffuse emission present within the FoV to guarantee a homogeneous background. Final values and their uncertainties are calculated as the mean and standard deviation of the Monte Carlo simulation results. In Section~\ref{subsec:non-param}, we present these indices and compare our results with the values obtained for the CEERS sample within $z$\,$\sim$\,0.8\,$-$\,9 \citep{Kartaltepe+23,Yao+23} and a sample of SMGs at $z$\,$<$\,5 \citep{Gillman+23}.

\subsection{Photometry and SED fitting}
\label{subsec:sed_fitting}

Along with the morphological analysis, we performed an SED fitting for all the galaxies in the sample. For this analysis, we extracted the photometry from the HST and MIRI images, covering the UV and near-IR rest-frame wavelengths from \simi0.1\,$\mu$m up to \simi3.6\,$\mu$m. In addition, we also used the ancillary photometric data available in the far-IR and (sub)-mm wavelengths, covering up to \simi650\micron rest-frame \citep{Liu+18}. 

The targets analysed in this work show different morphologies in the near-IR while in two of the DSFGs the rest-UV emission is very faint or does not correlate spatially with the rest near-IR (see Fig.~\ref{fig:UV-MIRI_images}). Therefore we extracted the photometry considering circular or elliptical apertures, centred on the F560W emission peak, large enough to fully cover both the UV and near-IR emission traced by the HST and MIRI filters, respectively. The typical sizes of these apertures are \simi 0.9\,$-$\,1.7\arcs of radius or semi-major axis (\simi 6\,$-$\,12\,kpc at these redshifts). Before extracting the fluxes, we performed a PSF match to make all the images have the spatial resolution of MIRI/F1280W, which has the largest FWHM within our sample (FWHM\,=\,0.42\arcsec), except for F1800W, where the detection of these galaxies is at a lower S/N (see Table~\ref{tab:fluxes}). 

During the photometry extraction, we considered GN20.2b, which shows a very concentrated brightness distribution, as a point-like source. Therefore, we applied an aperture correction using the empirical PSFs presented in \citet{Libralato+24}. This correction is \simi15\% for the circular aperture assumed ($r$\,=\,1.4\arcs). The standard deviation and level of the local background around our targets are determined for each image using annular apertures (1.6\arcsec<\,$r$\,<\,3.2\arcsec) centred on these objects, masking all the additional objects present in the FoV. We used these standard deviation values to estimate the associated uncertainty of the photometric measurements. These uncertainties were then re-scaled to take into account the correlated noise induced when drizzling the MIRI individual observations by multiplying them by a factor 2.24 (see \"Ostlin+\,in prep. for further details). For those filters that do not present significant emission, we adopted 3$\sigma$ upper limits. Table~\ref{tab:fluxes} presents the flux values measured from the HST and MIRI images along with the ancillary photometric points considered in the SED fitting. 

We used \texttt{CIGALE} \citep{Burgarella2005,Noll2009,Boquien2019} to perform the SED fitting. This is a Python-based modular code that allows us to create the SED model by defining separate modules for the different parameters (e.g. attenuation curves, dust emission, etc.). In this work, we assumed a star formation history (SFH) with continuous star formation across ages of between 1 and 20\Myr, along with a more mature stellar population created in an instantaneous burst of an age between 100\Myr and 1\,Gyr. For simplicity, we refer to these stellar populations as `young' and `mature' in the rest of the paper, respectively. We used the stellar population models from \citet{Bruzual&Charlot+03} with solar metallicity and a Chabrier initial mass function \citep[IMF,][]{Chabrier+03}. We included nebular emission with solar metallicity, an electron density of 100\,cm$^{-3}$, and an ionisation parameter of $\log(U)$\,=\,$-3$. Attenuation is assumed to follow the modified attenuation curve from \citet{Calzetti+00} presented by \citet{Noll2009} with the colour excess of the nebular lines and the slope from the power law values ranging between E(V$-$B)\,=\,[0.0, 2.1] and $\delta$\,=\,[$-$0.4, 0.6], respectively. A complementary analysis has shown that, with our data, we are not able to constrain the UV bump at 2175\,\AA. Besides, considering the bump does not modify our final conclusions, as the SED-fitting provide physical properties compatible within the uncertainties. Therefore, we fix the UV bump to be zero to reduce the number of free parameters. The IR emission was modelled using the \citet{Draine2014} dust models with PAH mass fraction and minimum radiation field values ranging between $q_\mathrm{PAH}$\,=\,[0.47, 2.5] and $U_\mathrm{min}$\,=\,[10, 50], respectively. The value of the power-law slope is set to $\alpha$\,=\,2.0 while the fraction of illumination is allow to vary between $\gamma$\,=\,[0, 0.02].


\section{Results and discussion}
\label{sec:4.Results}

\subsection{Near-IR structure and comparison with UV morphology}
\label{subsec:morph-results}


\begin{figure*}[!ht]
\centering
   \includegraphics[width=\linewidth]{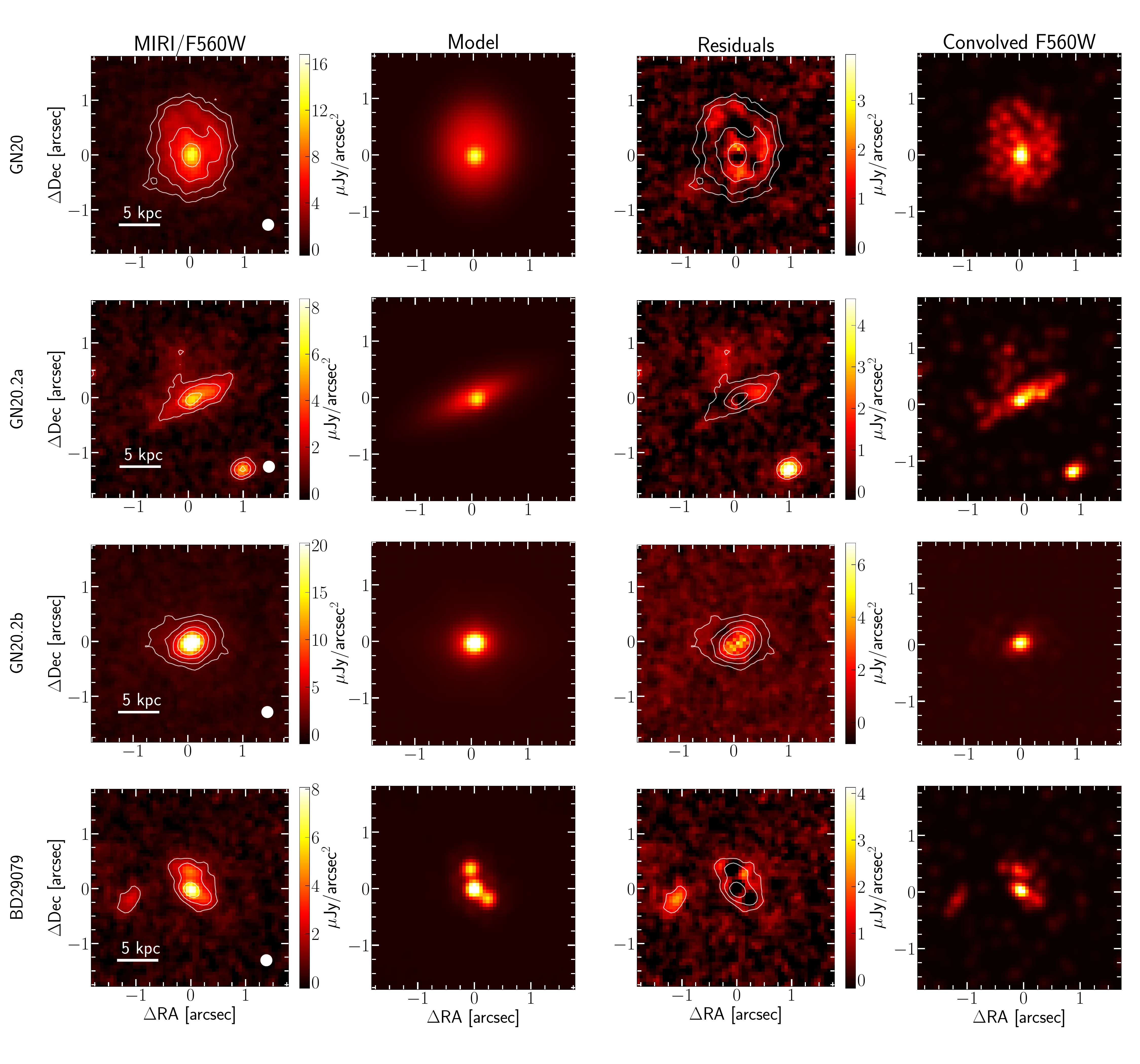}
      \caption{\texttt{Lenstronomy} modelling results for the sample. First, second, and third column display the original MIRI/F560W images, the \texttt{Lenstronomy} models and their residuals, respectively, for the objects of the sample. Fourth column shows the convolved F560W image obtained after applying a Lucy-Richardson deconvolution (see Sect.~\ref{subsec:Morph_analysis}).}
         \label{fig:lenstronomy}
\end{figure*}

\begin{table*}[]
    \centering
    \caption{S\'ersic results based on the MIRI/F560W images for the galaxies in the sample.}
\begin{tabular}{cccccc}
\hline \hline
  Galaxy &        $R_\mathrm{eff}$ &                     $n$ &                   $b/a$ &                         PA &  $i$ \\
 &       (kpc) &    &  &     (deg) &     (deg) \\
  \hline
GN20 &     3.51\pmm0.03 &  0.33\pmm0.02 &  0.80\pmm0.01 &  88\pmm2 &  38\pmm2\\
 GN20.2a &     2.13\pmm0.06 &  0.84\pmm0.10 &  0.25\pmm0.01 &  24\pmm1 &  81\pmm2 \\
 GN20.2b &     1.30\pmm0.05 &  4.51\pmm0.25 &  0.72\pmm0.03 &   6\pmm2 &  44\pmm2 \\
 \hline
 \end{tabular}
    \label{tab:Lenstronomy}
\end{table*}

In Figure~\ref{fig:UV-MIRI_images}, we present the HST and MIRI images for the target galaxies. At a first view, it is clear that there are differences between the rest-frame UV and near-IR emission in the DSFGs, while the LBG presents a similar morphology. The MIRI images of the DSFGs reveal, in general, extended centrally peaked brightness distributions that contrast with the compact and (in the case of GN20 and GN20.2a) off-centred UV emission.

The F1800W filter displays, in general, brightness distributions that are more diffuse and less centrally peaked than the bluer filters. This emission, which is more extended than the PSF size of this filter (see Fig.~\ref{fig:UV-MIRI_images}), suggests that the morphology at rest-frame \simi3.5\micron is different from the rest of the near-IR. At this redshift, the F1800W filter contains the PAH\,3.3\micron emission, which is typically strong in these star-forming objects \citep{Kim+12,Yamada+13}, as can be seen in the SED models from Sect.~\ref{subsec:SED-results}. Unlike the rest-frame UV emission traced by HST, the PAH emission, linked to the reemission of UV-photons produced by young stars, is extended and visible at the centre of these objects. This indicates that these galaxies are forming stars homogeneously throughout their extent and supports the scenario where the HST is only tracing the UV emission that leaks, in general, at the outskirts of these galaxies, far from the highly obscured central regions in the DSFGs. This scenario is also supported by the presence of molecular gas emission, traced by the PdBI CO(6–5) data presented by \citet{Carilli+10} and \citet{Hodge+13}, centred on the near-IR peak of the DSFGs (see Fig.~\ref{fig:UV-MIRI_images}). This CO emission indicates the existence of cold molecular gas, which fuel the star formation, available in the central regions of these galaxies, supporting the idea that there is much more star formation taking place than we are able to see in the rest-frame UV.

From the morphological modelling presented in Sect.~\ref{subsec:Morph_analysis}, we find, as in \citet{Colina+23}, that the MIRI/F560W image of GN20 can be represented by a compact nucleus (point-like) and an extended stellar envelope described with a S\'ersic profile with $n$\,=\,0.33\pmm0.02, $R_\mathrm{eff}$\,=\,3.51\pmm0.03\,kpc, and a high axial ratio of $b/a$\,=\,0.80\pmm0.01. This implies that if GN20 is a disc-like galaxy, assuming $q_0$\,=\,0.2, it has an inclination $i$\,=\,38\pmm2\,\degr, in agreement with the value derived from its CO(2-1) emission (i.e. $i$\,=\,30\pmm15\,\degr, \citealt{Hodge+12}). We also find an offset between the centroid of the extended stellar emission and the nucleus of \simi1\,kpc (see Fig.~\ref{fig:lenstronomy}), in agreement with \citet{Colina+23}. This offset is similar to those observed in local galaxies in clusters \citep{Lauer+88}, predicted to be the result of tidal interactions \citep{Aguilar&White+86}, suggesting that GN20 could have been part of a gravitational interaction. We obtained an aperture-corrected flux of 0.87\pmm0.32\uJy for the point-like source, whereas the flux for the extended component is 8.94\pmm0.32\uJy within the aperture used on the photometry extraction ($r$\,=\,1.4\arcs). These values are measured in the \texttt{Lenstronomy} model to avoid contamination between the different components,
representing \simi8\% and 83\% of the total flux, respectively; this is consistent with the values found by \citet{Colina+23}. The residuals from the modelling, happen to represent only a few percent of the original flux (i.e.  \simi9\%), displaying a patchy pattern, located higher at the outskirts of the extended stellar envelope. One of these bright residual regions (west blob) spatially coincides with the UV emission shown in the HST/F160W image (see Fig.~\ref{fig:UV-MIRI_images}). This agreement suggests spatial variations in the physical properties of the stellar envelope (e.g.  extinction), which allows the UV light to escape in the west part of the galaxy. A similar scenario has been found in HDF850.1, a DSFG embedded in an overdensity at $z$\,\simi5, where its UV and H$\alpha$ photons leak out far from its very obscured centre \citep{Herard-Demanche+23,Sun+24}.

The presence of a bright excess in the nuclear region of GN20.2a and its brightness asymmetry along the semi-major axis cause that this galaxy cannot be modelled with a single S\'ersic profile. We therefore also considered, as for GN20, the presence of a point-source component during the brightness distribution modelling. We find that after masking some diffuse emission present at the north-east of the object that is not confirmed to be part of GN20.2a, the extended emission from the galaxy can be modelled as a very elongated ($b/a$\,=\,0.25\pmm0.01) S\'ersic profile with $n$\,=\,0.84\pmm0.10 and $R_\mathrm{eff}$\,=\,2.13\pmm0.06\,kpc. The axial ratio of the galaxy implies, assuming that GN20.2a is a disc-like galaxy with $q_0$\,=\,0.2, an inclination of $i$\,=\,81\pmm2\,\degr. Contrary to GN20, we do not find any offset between the S\'ersic and the point-like components, suggesting that this source is likely linked to the nucleus of the galaxy. The fluxes of the extended and point sources are 1.98\pmm0.23\uJy and 0.29\pmm0.23\uJy (after aperture correction), respectively, which represent 60\% and 9\% of the total flux. In this galaxy, the residuals represent about 31\% of original flux, which can be understood as due to the diffuse emission in the NE of the galaxy visible in HST/F160W and MIRI/F560W (see Fig.~\ref{fig:UV-MIRI_images}) and to light excess along the NW semi-major axis. In fact, this asymmetry along the major axis, depicted by the contours in Figure~\ref{fig:lenstronomy}, can be also observed in the residual map. Possible origins of this asymmetry could be an irregular dust distribution or/and the presence of very bright blobs located somewhere in the NW semi-major axis, intensified by the high inclination of the galaxy. The Lucy-Richardson convolved MIRI image (rightmost panel in Fig.~\ref{fig:lenstronomy}) supports this last scenario, showing some bright blobs along the NW semi-major axis. Unlike the previous object, GN20.2a does not present any bright HST/F160W counterpart (see Fig.~\ref{fig:UV-MIRI_images}) while showing some diffuse emission at \simi3\,kpc to the NE of the galaxy, spatially coincident with the extended emission seen in MIRI/F560W. Further photometric analysis of this emission is needed to determine if it is really part of GN20.2a.

GN20.2b shows clear differences in the morphology with respect to the other two DSFGs. It presents a compact brightness distribution that can be modelled by a single S\'ersic profile with $n$\,=\,4.51\pmm0.25, $R_\mathrm{eff}$\,=\,1.30\pmm0.05\,kpc, and $b/a$\,=\,0.72\pmm0.03. We have tested, for completeness, the possibility of the presence of a point-source along with the S\'ersic profile. However, we discarded this approach as it yields larger residuals and a higher Bayesian information criterion (BIC) index. The residuals of the single S\'ersic profile display an elongated (to the NW) structure, although it might be created by two close intense blobs. These residuals are aligned with the slightly elongated inner contours from the original F560W image (see Fig.~\ref{fig:lenstronomy}), which disagree with the external orientation depicted with the outermost contours (PA\,\simi0\,\degr). This difference in the orientation of the inner regions may indicate the presence of an internal sub-structure, such as an inner disc. Another possible scenario is that GN20.2b is an object in a merging stage very close to the post-coalescence, in which the gravitational interaction could have created this difference between the inner and outer isophotes. Further observations of this target at a better spatial resolution and larger exposure time are needed to shed light to its peculiar morphology. The rest-frame UV image of this galaxy (see Fig.~\ref{fig:UV-MIRI_images}) presents a compact intense emission in the very central region of the galaxy, which is spatially coincident with the near-IR brightness peak. This result contrasts with the other two DSFGs, where we do not find any emission in the nuclear regions, suggesting that this object could be less extinguished in its nucleus.

The visual inspection of the Lyman-break galaxy (i.e.  BD29079) reveals that the stellar component of this object is composed of three bright clumps very close to each other (i.e.  $r$\,$<$\,4\,kpc). Due to its complex sub-structure, we decided to use a different approach to model its morphology. We fit its brightness distribution considering three independent point-sources, allowing them to vary their intensity. With this approach, we found that the central, north and southeast clumps present fluxes of 1.26\pmm0.15\uJy, 0.71\pmm0.15\uJy, and 0.52\pmm0.15\uJy, respectively. The residual map from this model, shown in Fig.~\ref{fig:lenstronomy}, displays a brightness excess around the central and north blobs. This suggests that these blobs might be more extended than the PSF or, at least, very close to the resolution limit. Therefore, we tried to model these three blobs simultaneously with S\'ersic profiles, fixing their centres based on the point-source results. However, the $\chi^2$ and BIC values of this new approach are worse than those obtained from the three point-source model. We attribute this result to the fact that, although these regions are probably partially-resolved, their separation (\simi1.5\,$\times$\,FWHM) makes it difficult to correctly model them as independent S\'ersic profiles. Therefore, we assume the three point-source model as the best approach for our data. For comparison with the DSFGs sizes, which are modelled with S\'ersic profiles, we have derived the effective radius of BD29079 using the curve-of-growth (CoW) method. We find that this galaxy, although clumpy, is dominated by its central emission with a $R_\mathrm{eff}$\,=\,1.52\pmm0.05\,kpc. When we compare the F560W with the UV image, we observe that unlike the DSFGs, this galaxy shows a very similar structure in HST/F160W (see Fig.~\ref{fig:UV-MIRI_images}), with a central bright clump and some diffuse emission spatially coincident with the near-IR clumps. This match between the rest-frame UV and near-IR structure might be produced by an intense UV emission coming from massive young stars, typical in these objects \citep[see][for a review]{Giavalisco+02} along with a lower amount of dust that would make this object less extinguished in its nuclear regions.


We repeated the same procedure with the MIRI/F770W images (i.e.  H-band rest-frame) for all the galaxies. In general, we find a good agreement (with differences $<$\,15\%) in the effective radii and S\'ersic indices to those obtained from the F560W images (i.e.  J-band). This result is expected since the near-IR emission traces a similar mature stellar population dominated by cool supergiant stars with some contribution of the asymptotic and red giant branches (i.e.  AGB and RGB; \citealt{Mouhcine+02,Valenti+04,Maraston+05}). In addition, this agreement also implies that possible uncertainties in the PSF-shape determination and spatial resolution difference are not decisive in our morphological modelling.

The typical sizes of our DSFGs (i.e.  $R_\mathrm{eff}$\,\simi1.3\,$-$\,3.5\,kpc) are up to about two times larger than what is expected from the size-decreasing trend observed by \citet{Gillman+23} for SMGs at 1$<$\,$z$\,$<$\,4 using JWST rest-frame near-IR images. In fact, our objects are comparable in size with the SMGs at $z$\,\simi2 analysed in that work. This difference in size is specially noticeable in GN20 and GN20.2a, which show disc-like morphologies. Their higher star formation rates (SFR\,$>$\,1000\Msyr, see Sect.~\ref{subsec:SED-results}),  molecular gas reservoirs ($M_\mathrm{gas}$\,>\,2\EXP{11}\Ms, \citealt{Tan+14}), and molecular gas fractions ($f_\mathrm{gas}$\,$>$\,0.65, \citealt{Tan+14}) might play a role in their size evolution and morphology; this scenario  stands in contrast to the lower gas content ($f_\mathrm{gas}$\,$<$\,0.45, \citealt{Tan+14}) and more compact ($R_\mathrm{eff}$\,=\,1.30\pmm0.05\,kpc) structure of GN20.2b, which is more similar in size to the $z$\,$>$\,3 SMGs analysed in \citet{Gillman+23}. Regarding the S\'ersic index and axial-ratios, our sample shows a wide range of values ($n$\,\simi0.3$-$4.5, $b/a$\,\simi0.3\,$-$\,0.8), compatible with the typical results obtained for SMG at $z$\,$>$\,2 \citep{Gillman+23}. According to the results drawn from CEERS near-IRCam imaging for a general sample of galaxies \citep{Kartaltepe+23}, at $z$\,\simi4, the typical size for disc-like, spheroidal, and irregular galaxies ranges between $R_\mathrm{eff}$\,\simi0.5\,$-$\,1.5\,kpc. Therefore, our results show  that GN20 and GN20.2a are up to three times larger than the general isolated galaxies at their redshift, whereas GN20.2b presents a similar size.

In summary, while the DSFGs from the overdensity show S\'ersic indices and axial-ratios compatible with the typical values observed in similar galaxies at their redshift, our targets are larger in terms of effective radii (especially GN20). This could be explained as a rapid growth  in size due to their membership in the overdensity. In Sect.~\ref{subsec:comparison_local_OD}, we discuss how these galaxies compare in mass and SFR with the general population at $z$\,\simi4.

\subsection{Non-parametric morphological indices}
\label{subsec:non-param}

\begin{figure*}[!ht]
\centering
   \includegraphics[width=\linewidth]{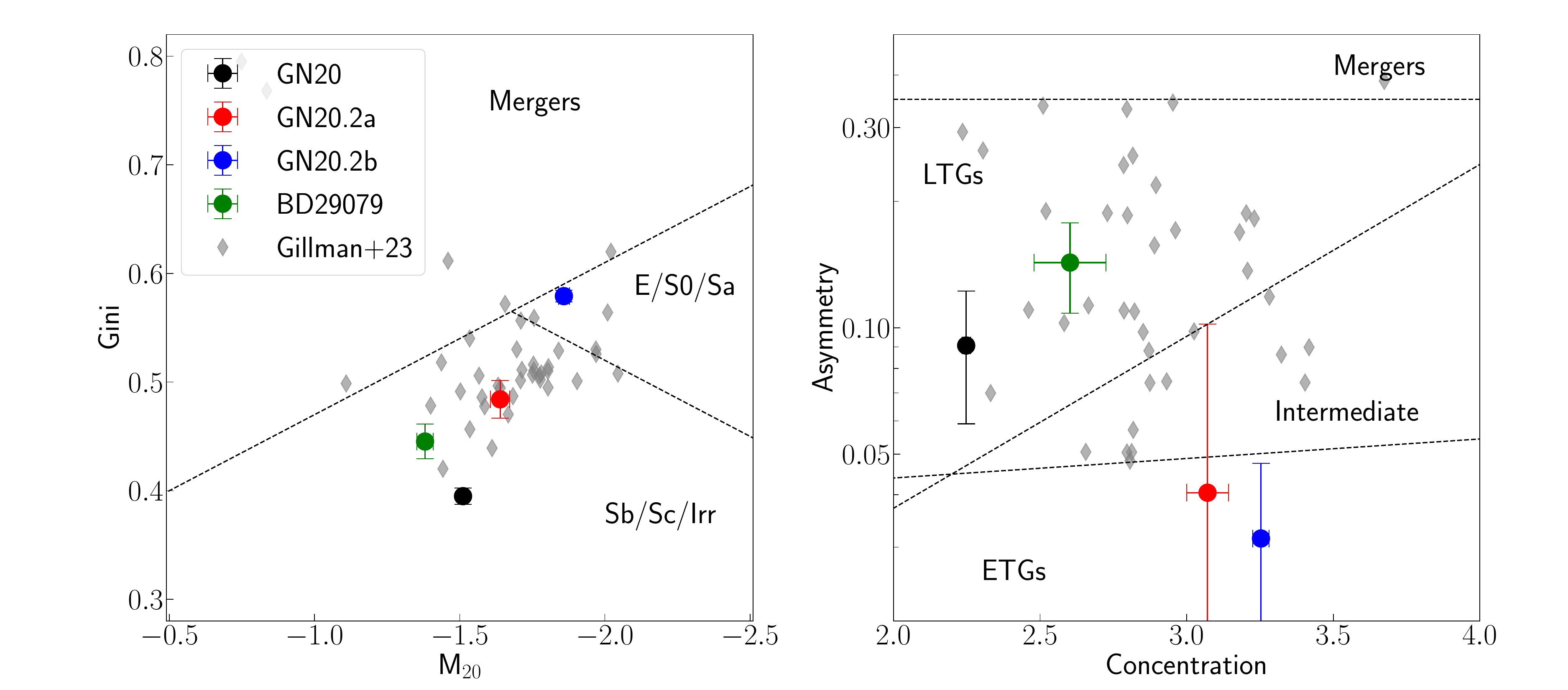}
      \caption{Morphological classification based on non-parametric indices. Gini\,$-$\,$M_{20}$ (left) and $C$\,$-$\,$A$ (right) panels used to classify the morphology for objects in our sample. Dashed lines separate between ETGs, LTGs, and merging systems following \citet{Bershady+00}, \citet{Conselice+03}, and \citet{Lotz+08a}. Grey diamonds represent the sample of SMGs observed at rest-frame near-IR by \citet{Gillman+23}.}
         \label{fig:CAS}
\end{figure*}

\begin{table*}[]
    \centering
    \caption{MIRI/F560W non-parametric indices for the galaxies in the sample.}
\begin{tabular}{cccccc}
\hline \hline
  Galaxy & $C$ & $A$ & $S$ & Gini & $M_\mathrm{20}$ \\
\hline
    GN20 &      2.25$\pm$0.01 &      0.09$\pm$0.03 &      0.02$\pm$0.01 &      0.39$\pm$0.01 &       -1.51$\pm$0.02 \\
   GN20.2a &      3.07$\pm$0.07 &      0.04$\pm$0.06 &     -0.04$\pm$0.08 &      0.48$\pm$0.02 &       -1.64$\pm$0.03 \\
   GN20.2b &      3.25$\pm$0.03 &      0.03$\pm$0.02 &     -0.00$\pm$0.01 &      0.58$\pm$0.01 &       -1.86$\pm$0.02 \\
 BD29079 &      2.60$\pm$0.12 &      0.14$\pm$0.03 &      0.03$\pm$0.03 &      0.45$\pm$0.02 &       -1.38$\pm$0.03 \\
 \hline
 \end{tabular}
    \label{tab:Statmorph}
\end{table*}

As introduced in Section~\ref{subsec:Morph_analysis}, we used \texttt{Statmorph} to derive non-parametric image statistics for the MIRI filters with the best spatial resolution (i.e.  F560W and F770W). This code allows us to compute the Concentration ($C$), Asymmetry ($A)$ and Smoothness/Clumpiness ($S$) parameters (i.e.  $CAS$ indices) along with the Gini ($G$) and $M_{20}$ values \citep{Bershady+00,Conselice+03,Abraham+03,Lotz+04}. Basically, higher values for the $C$, $A$ and $S$ indices indicate more concentrated, more asymmetric and clumpier surface brightness profiles, respectively. The Gini coefficient represents the pixel distribution of the galaxy brightness, where $G$\,=\,0 indicates a homogeneous light distribution among all the pixels while $G$\,=\,1 represents the extreme case in which all the light is concentrated in a single pixel. The $M_{20}$ parameter is defined as the normalised
second-order moment of the brightest 20\% of the pixels, regardless of whether they are in the central region or not; thus, a high concentration of light will result in very negative $M_{20}$ values. Table~\ref{tab:Statmorph} presents the non-parametric indices for the MIRI/F560W images for  the whole sample. 

These non-parametric indices have been commonly used to define the morphological classification, distinguishing between mergers, disc-like and elliptical galaxies \citep[e.g.][]{Bershady+00,Conselice+03,Lotz+08a}. Left panel of Fig.~\ref{fig:CAS} shows the 
Gini-$M_{20}$ plane where the dashed lines mark the boundaries between early- (ETGs), late-type galaxies (LTGs) and mergers as defined by \cite{Lotz+08a}. Although all the objects in the sample present similar $M_{20}$ values, the lower Gini values for GN20, GN20.2a, and BD29079, indicating less concentrated brightness distributions, classify them as late-type galaxies, whereas GN20.2b falls into the early-type galaxy region. This agrees with the visual inspection and the parametric modelling performed in Sect.~\ref{subsec:morph-results}, which show GN20.2b as a compact symmetrical object with high S\'ersic index (i.e.  bulge-like); whereas GN20 and GN20.2a present an extended disc-like emission with the presence of some brightness asymmetries. When looking at the concentration-asymmetry plane (right panel, Fig.~\ref{fig:CAS}), we obtained similar results. While GN20 and BD29079 are in the LTGs region, GN20.2b is classified as an early-type galaxy. Finally GN20.2a is identified as an ETG, though the uncertainty in the asymmetry makes it compatible with an LTG or an intermediate object. None of the galaxies are classified as merging objects in either panel. For the LBG BD29079, which displays a very clumpy structure in both rest-frame UV and near-IR wavelengths, we obtained a relatively high asymmetric value ($A_\mathrm{F560W}$\,=\,0.14\pmm0.03), but a low clumpiness index (i.e.  $S$\,=\,0.03\pmm0.03). This could be an effect of the spatial resolution, which would smooth the internal sub-structure and flatten the brightness profile, as it has been also observed for a larger sample of SMGs at $z$\,$<$\,5 \citep{Gillman+23}. In fact, in \citet{Yao+23} the authors found that, for a sample of \simi1300 galaxies drawn from the CEERS field at $z$\,$<$\,3, larger PSF FWHMs tend to decrease the clumpiness and increase the Gini and $M_{20}$ indices. 

We  also computed these non-parametric indices based on the F770W images, whereas the F1280W and F1800W cutouts were discarded for this analysis due to their lower S/N and worse spatial resolution (FWHM\,$>$\,0.42\arcs). For the F770W images, we obtained similar results to those from F560W, with typical differences $<$\,10\% and, in general, compatible within the uncertainties. These small differences do not modify the conclusions derived from the non-parametric analysis.


Therefore, based on non-parametric indices, our finding that three out of our four galaxies at $z$\,\simi4 are classified as LTGs is compatible with the result found for a sample of more than 40 SMGs at $z$\,$<$\,5 \citep{Gillman+23}. A similar result was also found for a sample of $>$1000 galaxies from the CEERS sample at lower redshifts ($z$\,$<$\,3, \citealt{Yao+23}). In a complementary way, \citet{Kartaltepe+23} analyse the non-parametric indices for $>$800 galaxies in the CEERS field at 3\,$<$\,$z$\,$<$\,9. The galaxies from that work happen to be predominantly disc-like, especially up to $z$\,\simi5, where they represent \simi40\% of the total sample. In this context, it is not surprising to see that the three out of the four galaxies in the overdensity at $z$\,\simi4 studied in this work are classified as LTGs.

\subsection{SED fitting results}
\label{subsec:SED-results}


\begin{figure*}[!ht]
\centering
   \includegraphics[width=\linewidth]{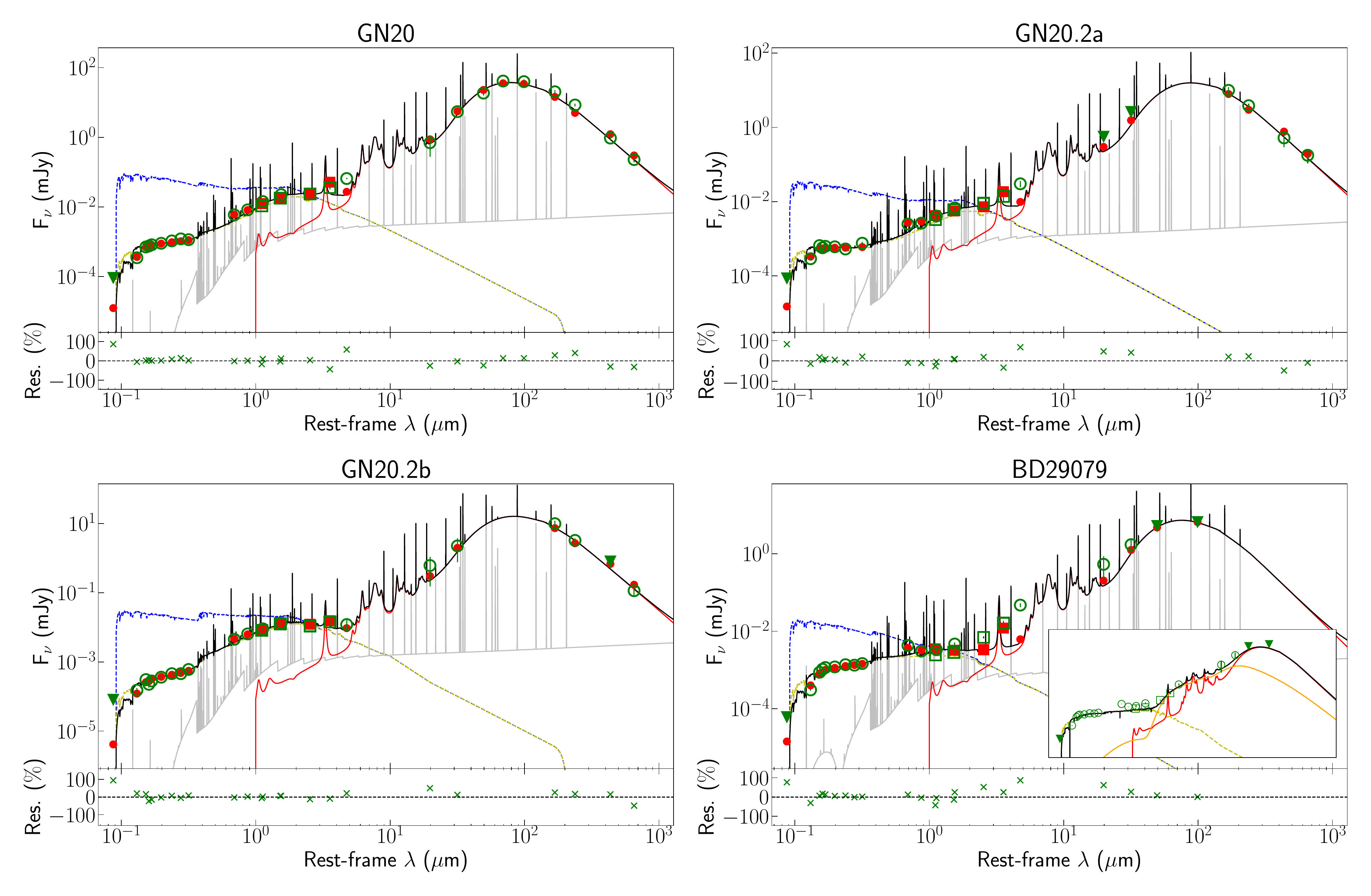}
      \caption{Rest-frame UV to millimetre (mm) SEDs and best-fit model derived with \texttt{CIGALE} for the objects in the sample. Green open circles and triangles show the observed fluxes and upper-limits as presented in Table~\ref{tab:fluxes}, while the MIRI photometric measurements are marked as green open-squares. Lines represent the stellar (unattenuated in dashed blue and attenuated in dashed yellow), nebular (solid grey), and dust (in solid red) contribution to the total SED (in black). Red circles display the modelled fluxes in every filter considered during the SED fitting. Each panel shows, at the bottom, the relative residuals derived from the measured photometry and the \texttt{CIGALE} results. Orange solid line in the inset displays the AGN contribution considered for BD29079 (see Sect.~\ref{subsec:SED-results}). In this inset, the nebular contribution is not plotted for simplicity.}
         \label{fig:SED}
\end{figure*}

\begin{table*}[]
    \centering
    \caption{SED fitting results.}
    \begin{tabular}{cccccccc}
    \hline \hline
Galaxy &      \Mstar$^\mathrm{tot}$ & $M_\mathrm{*}^\mathrm{young}$ &    SFR & sSFR  & Age$_\mathrm{young}$ & $\delta$ & $A_\mathrm{V}$  \\
 &     (10$^{10}$\Ms) & (10$^{9}$\Ms) &  ($M_\odot$\,yr$^{-1}$) &  (Gyr$^{-1}$)  & (Myr) &  & (mag) \\
    \hline 
   GN20 &   8.6$\pm$4.3 &            45.0$\pm$5.2 &  2550$\pm$150 &  30$\pm$15 &         18.0$\pm$3.0 &   0.29$\pm$0.03 &  1.50$\pm$0.01 \\
   GN20.2a &   2.2$\pm$1.4 &             13.4$\pm$2.3 &  1030$\pm$130 &  47$\pm$32 &         14.3$\pm$2.8 &    0.40$\pm$0.05 &  1.22$\pm$0.05 \\
   GN20.2b &  17.9$\pm$3.9 &             8.4$\pm$3.4 &   760$\pm$230 &    4.2$\pm$1.6 &          8.5$\pm$5.3  &   0.19$\pm$0.07 &  1.48$\pm$0.09 \\
 BD29079 &   2.4$\pm$0.7 &             3.1$\pm$0.7 &    310$\pm$80 &   13.0$\pm$5.0 &          2.4$\pm$1.9  &  -0.36$\pm$0.12 &  0.71$\pm$0.08 \\
 \hline 
    \end{tabular}
    \label{tab:SED_results}
     \tablefoot{The `young' stellar mass corresponds to the stellar population created during the last 20\Myr (see Sect.~\ref{subsec:sed_fitting}). The SFR is referred to the last 10\,Myr, linked to this young stellar population. The sSFR is derived using the SFR and total stellar mass resulting from the SED fitting.}
\end{table*}

The SEDs of the galaxies in our sample have been previously studied based on existing UV to FIR rest-frame photometric points \citep{Tan+13,Tan+14}. With the new MIRI observations, we are able to cover the near-IR range up to \simi3.5\micron, helping us to better constrain the mass and age of a possible mature stellar population, found previously in DSFGs \citep{Ma+15}. As introduced in Sect.~\ref{subsec:sed_fitting}, we used \texttt{CIGALE} to fit the rest-frame UV-to-mm SEDs for the objects in our sample. Figure~\ref{fig:SED} shows the best-fit SEDs obtained while Table~\ref{tab:SED_results} summarises the physical parameters derived from the Bayesian analysis carried out in this work (see Sect.~\ref{subsec:sed_fitting}).

For GN20, we found that the young (18\pmm3\Myr) stellar component represents \simi52\,\% of the total stellar mass of the galaxy ($M_\mathrm{*}^\mathrm{tot}$\,=\,(8.6\pmm4.3)\EXP{10}\Ms). This dominance of the young population can be understand due to the high SFR derived for this object (i.e. \,2550\,\pmm150\Msyr). The global extinction, in the V-band, of this object ($A_\mathrm{V}$\,=\,1.50\pmm0.01\,mag) is expected to be even greater in the nuclear region, where we do not observe rest-frame UV emission. This value is smaller than the extinctions derived by \citet{Bik+24} based on the difference between the SFR$_\mathrm{IR}$ and SFR$_\mathrm{Pa\alpha}$ and assuming either a Cardelli extinction law (i.e.  $A_\mathrm{V}$\,=\,17.2\,mag; \citealt{Cardelli+89}) or the star+gas mixed model (i.e.  $A_\mathrm{V,mixed}$\,=\,44\,mag) presented in \citet{Calabro+18}. The large difference between our SED-fitting $A_\mathrm{V}$ and the one derived based on Pa$\alpha$ might be understood as the presence of a stellar population older than \simi10\,Myr that still forms stars and contributes to the total SFR but that cannot be traced by the Pa$\alpha$ emission. Another potential origin of this difference is the possibility that GN20 contains a buried AGN, as proposed by \citet{Riechers+13}, that would enhance the total SFR but is not traced by Pa$\alpha$. 

The SED fitting revealed that GN20.2a is the least massive DSFG in the sample, with $M_\mathrm{*}$\,=\,(2.2\,$\pm$\,1.4)\,$\times$\,10$^{10}$\,$M_\odot$, producing stars at a rate of 1030\pmm130\Msyr. In this object, the young stellar population represents a larger fraction of the total mass (\simi59\,\%) than in GN20. GN20.2b is the galaxy among the DSFGs with the lowest SFR (i.e.  760\pmm230\Msyr). This massive galaxy (\Mstar=\,(1.8\pmm0.4)\EXP{11}\Ms) presents a young burst (8.5\pmm5.3\Myr), likely linked to the bright central UV source visible in the HST (see Fig.~\ref{fig:UV-MIRI_images}). This burst is creating a young stellar population that represents only \simi5\,\% of the total mass, while the bulk of the stellar mass of this galaxy is dominated by a more mature stellar population. The SED model shows (see Fig.~\ref{fig:SED}) that this galaxy presents the larger 4000\AA\xspace break in the sample, in line with its ETG classification during the morphology analysis (see Sect.~\ref{subsec:morph-results}). The dominance of the mature stellar population in the stellar mass budget along with the young age of the burst seems to suggest that this galaxy is experiencing a "rejuvenation" process, as observed in some LBGs and LAEs at similar redshifts \citep{Angthopo+19,Rosani+20,Iani+24}.

For the Lyman-Break galaxy BD29079, we found that it presents a total stellar mass of (2.4\pmm0.7)\EXP{10}\Ms, similarly to GN20.2a, while the very young (2.4\pmm1.9\Myr) population represents \simi15\% of the total stellar mass budget. The presence of this very young population along with the low extinction ($A_\mathrm{V}$\,=0.71\pmm0.08\,mag) of this galaxy might explain the strong emission observed in the rest-frame UV (see Fig.~\ref{fig:UV-MIRI_images}). Figure~\ref{fig:SED} shows that there are some residuals in the mid-IR due to the presence of a power-law type behaviour that is not captured by the \texttt{CIGALE} model for this galaxy. This feature is linked to the presence of some warm dust component, usually produced by an AGN. Therefore, for this galaxy, we tested the possibility of modelling the AGN contribution using the \citet{Fritz+06} models available in \texttt{CIGALE} following the prescriptions presented in \citet{Ciesla+15}. This new SED model, displayed in the inset in Fig.~\ref{fig:SED}, improves the $\chi^2$ value and reduces the residuals in the mid-IR. Under this approach, the AGN would contribute up to a \simi20\% to the total $L_\mathrm{IR}$.

Considering a modified Calzetti extinction law (see \citealt{Noll2009}) during the SED fitting allows us to study the possible variation in the dust attenuation curve for these galaxies. We found that the DSFGs present positive values of the power-law slope ($\delta$\,\simi0.2\,$-$\,0.4), indicating an extinction law shallower than Calzetti's, while the analysis on the LBG yields a steeper curve ($\delta$\,=\,$-$0.36\pmm0.12). This agrees with the statistical analysis performed in $>$10\,000 LBGs at z\,\simi3 presented by \citet{Alvarez-Marquez+19}, where dust attenuation laws steeper than Calzetti's are found to be favoured in LBGs with log(\Mstar/\Ms)\,<\,10.3. The positive $\delta$ values derived for the DSFGs are consistent with the general trend in which shallower attenuation curves are found in more massive, dust-obscured star-forming galaxies \citep{Salim+18,Salim+20}.

In general, our results derived from the rest-frame UV to millimetre SED fitting (e.g.  \Mstar, SFR) are in good agreement with the values obtained in previous studies. In \citet{Tan+14}, the authors performed an SED analysis with \texttt{MAGPHYS}, based on UV to far-IR photometry, for the four galaxies in the overdensity. They derived total stellar masses equal to 1.1\EXP{11}, 3.8\EXP{10} and 1.1\EXP{11}\Ms for GN20, GN20.2a, and GN20.2b respectively, with a typical error of 0.2\,dex. While compatible within the 1$\sigma$ uncertainties, our stellar masses are slightly lower, except for GN20.2b for which we obtain a stellar mass \simi1.5 times larger (see Table~\ref{tab:SED_results}). The stellar mass derived via SED for the LBG is also in good agreement with the result obtained by \citet{Tan+13}, namely,  (2.5\pmm0.5)\EXP{10}\Ms. Regarding the SFRs, our results for the DSFGs agree with the values derived by \citet{Tan+14} with typical differences $<$\,20\%, while our typical uncertainty on the SFR is \simi10\%. For BD29079, we also obtained an SFR value compatible within the 1$\sigma$ uncertainties with previous estimations presented by \citet{Tan+13}. This general agreement in both the stellar mass and SFR makes that our specific SFR (sSFR) displays similar values, ranging between \simi4\,$-$\,50\,Gyr$^{-1}$. When comparing with previous estimations of the dynamical mass, we observe that our stellar mass value for GN20 is a factor of \simi2\,$-$\,6 smaller than the dynamical masses derived by \citet{Tan+14} and \citet{Hodge+12} (i.e.  (1.8\,$-$\,5.4)\EXP{11}\Ms), based on its gas+stellar mass and its CO kinematics, respectively. This difference is even larger in the case of GN20.2a, where the stellar mass derived in this work is a factor \simi10 smaller than the $M_\mathrm{dyn}$ presented in \citet{Tan+14} (i.e.  (1.9\pmm0.9)\EXP{11}\Ms).

\subsection{Comparison with isolated SFGs and similar overdensities}
\label{subsec:comparison_local_OD}

\begin{figure}[!ht]
\centering
   \includegraphics[width=\linewidth]{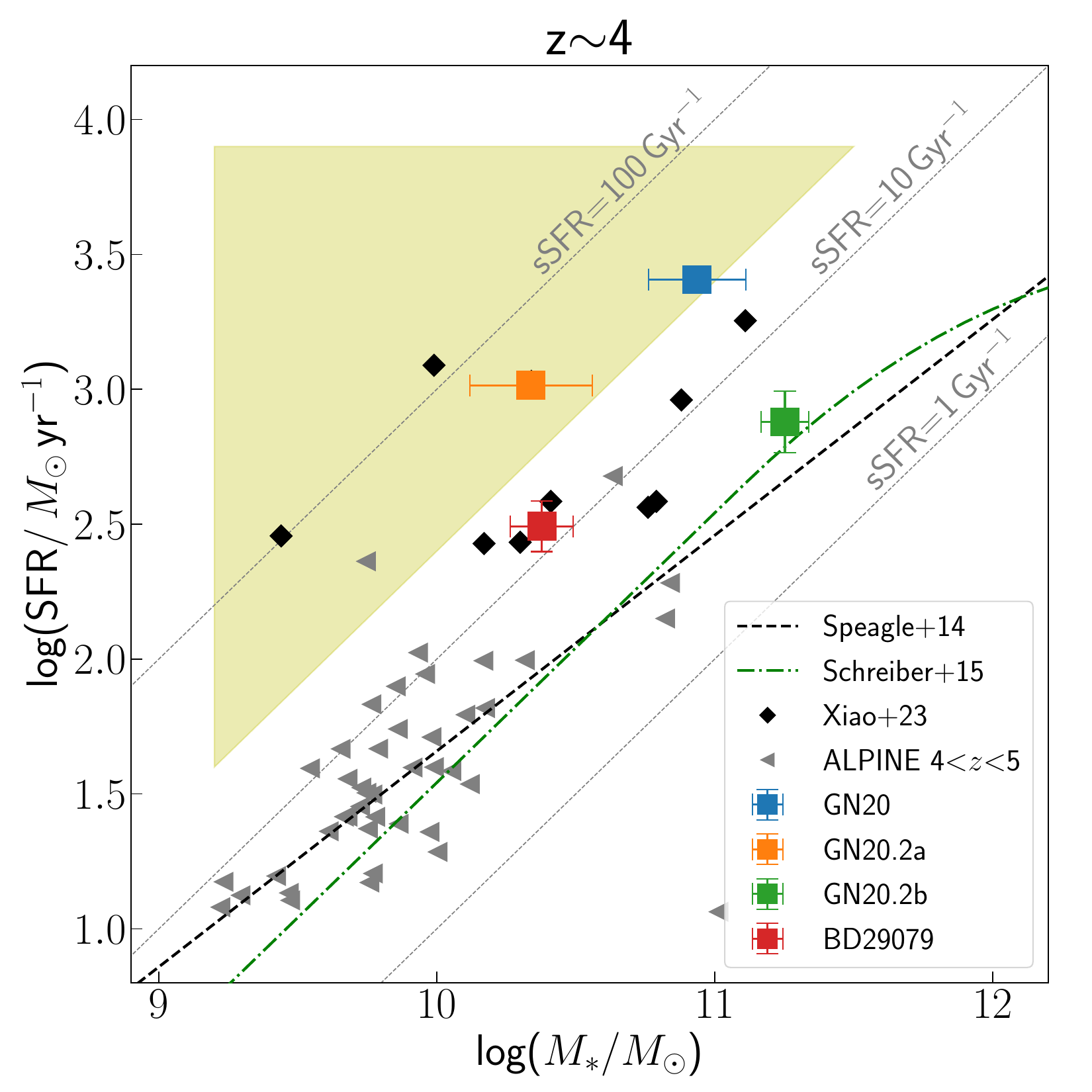}
      \caption{SFR versus stellar mass plane. The panel shows the galaxies studied in this work, as squares, along with the ALPINE sample at similar redshift \citep{Faisst+20}, as grey triangles, and the sample of OFGs from \citet{Xiao+23}, as black diamonds. Black dashed and green dot-dashed lines represent the SFMS at $z$\,=\,4 following \citet{Speagle+14} and \citet{Schreiber+15}, respectively. Yellow shadow area represents the starburst region defined by \citet{Caputi+17}.}
         \label{fig:SFR_Ms}
\end{figure}

\begin{figure*}[!ht]
\centering
   \includegraphics[width=\linewidth]{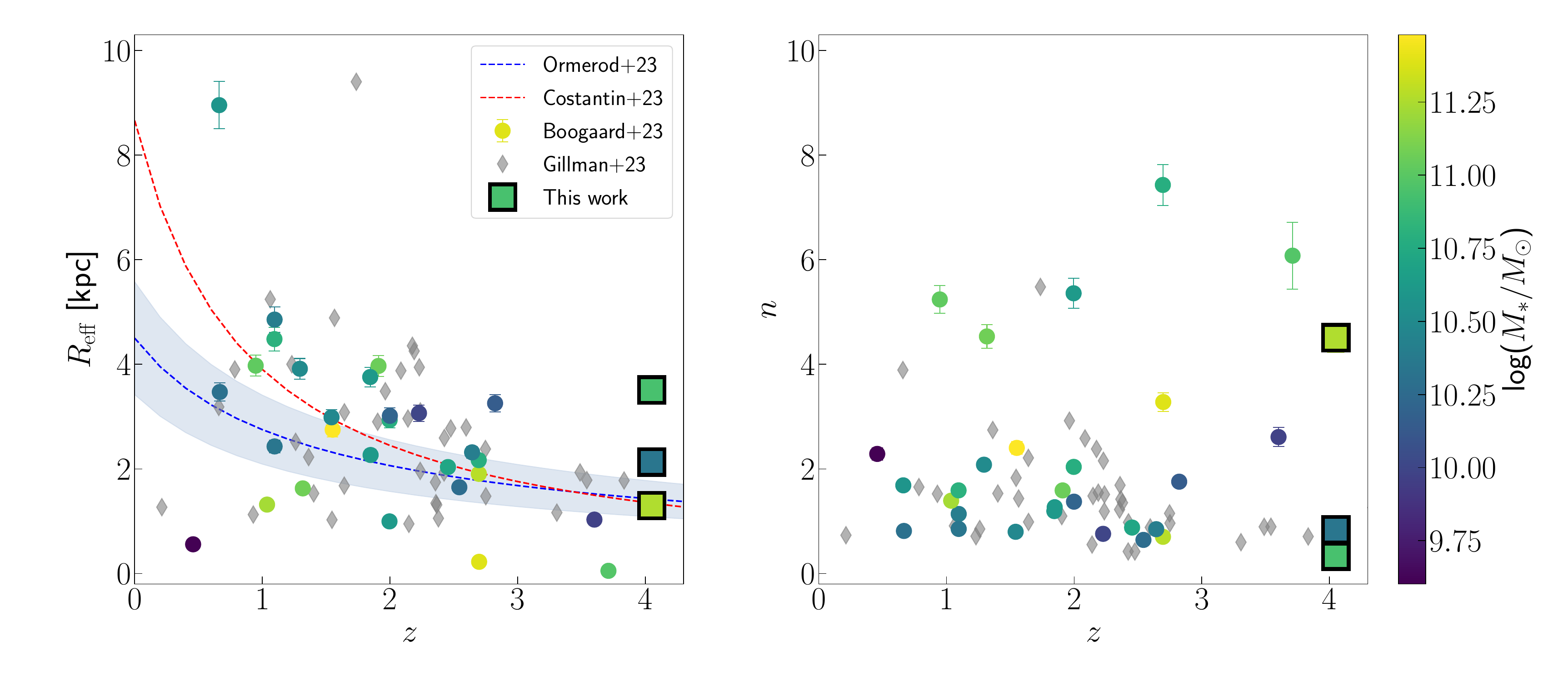}
      \caption{Evolution of morphology with redshift. This figure shows the evolution with redshift for the effective radii (left panel) and S\'ersic index (right panel) for the sample of ALMA-selected galaxies presented in \citet{Boogaard+24}. Black-edged squares display the position of our DSFGs in these panels. Each point is colour-coded according to its stellar mass. Red dashed line in the left panel represents the trend found in \citet{Costantin+23_TNG} for CEERS F356W mock images produced from TNG50 simulations. The blue line and shadow show the best-fit and its associated uncertainty derived for a sample of CEERS+CANDELS galaxies by \citet{Ormerod+24}. Grey diamonds display the sample of SMGs at $z$\,<\,5 presented by \citet{Gillman+23}.}
         \label{fig:z_evol}
\end{figure*}

The morphological and SED-fitting analyses performed in this work have revealed that despite displaying a wide range of morphologies, all the objects from our sample  are extended and massive star-forming galaxies overall. When plotted on the SFR$-$\Mstar plane (Fig.~\ref{fig:SFR_Ms}), GN20, GN20.2a and BD29079 lie above the star-forming main sequence (SFMS) at $z$\,\simi4 \citep{Speagle+14,Schreiber+15} by more than the intrinsic scatter of the MS (i.e.  \simi0.3\,dex,\,\citealt{Whitaker+12}), while GN20.2b is compatible with being in high-mass end of the main sequence. This is consistent with the morphological results, which revealed that GN20.2b is the only object in the sample classified as ETG, presenting the highest S\'ersic index and the smallest effective radius. This difference in the sSFR (up to a factor of \simi10) suggests that these galaxies are in different evolutionary stages. While GN20b is at an advanced stage (i.e. lower sSFR, larger mass and more compact structure), GN20 and GN20.2a are in a earlier stage, forming stars at a larger rate. Although GN20, GN20.2a and BD29079 are above the SFMS, only the two DSFGs present sSFR compatible with the "Starburst region" defined in \citet{Caputi+17}, namely, \, log(sSFR/yr$^{-1}$)\,>\,$-$7.6. When making comparisons with other galaxies at similar redshifts, we generally observe that our sample is almost 1\,dex more massive than the [CII] emitters from ALPINE at similar redshifts \citep{Faisst+20}, while it is more similar to the optically-faint galaxies (OFGs) at $z$\,\simi4 drawn from the GOODS-ALMA sample \citep{Xiao+23}. 

As introduced above, previous CO-based analyses have calculated the gas mass of these galaxies \citep{Carilli+11,Hodge+13,Tan+13,Tan+14}.  Depending on the $\alpha_\mathrm{CO}$ assumed, these masses
range between (1\,$-$\,2)\EXP{11}\Ms, (0.6\,$-$\,2)\EXP{11}\Ms and (2\,$-$\,8)\EXP{10}\Ms for GN20, GN20.2a and GN20.2b, respectively. These gas masses imply a gas-to-stellar mass ratio ($\mu_\mathrm{g}$\,=\,$M_\mathrm{g}$/\Mstar{}) of \simi1\,$-$\,3 for GN20 and GN20.2a, whereas in GN20.2b the stellar mass dominates (i.e. \,$\mu_\mathrm{g}$\,\simi0.1\,$-$\,0.4). The fact that GN20 and GN20.2a are more gas-rich supports the scenario in which GN20.2b is in a more advanced stage of galaxy evolution. Based on the SED-derived SFR, we obtain very short depletion times ($\tau$\,=\,$M_\mathrm{g}$/SFR\,\simi30\,$-$\,100\,Myr) for the three DSFGs. This is consistent with the depletion times found in other DSFGs, both isolated and in overdense regions, at different redshifts  \citep{Bethermin+15,Aravena+16,Gururajan+22,Zavala+22}.

Since the launch of the JWST, several studies have used the spatial resolution and sensitivity of its cameras to analyse the size evolution up to $z$\,\simi8 for different samples of observed (e.g.  CEERS, \citealt{Gomez-Guijarro+23,Ormerod+24,Ward+24}; ALMA-HUDF, \citealt{Boogaard+24}) and simulated galaxies \citep{Costantin+23_TNG}. These works have found that, at higher redshifts, galaxies tend to be more compact, regardless of their stellar mass or the wavelength range studied. Figure~\ref{fig:z_evol} (left panel) displays the size-evolution of ALMA-selected sources from the Hubble Ultra Deep Field (HUDF) observed with MIRI/F560W \citep{Boogaard+24}. It is clear that the DSFGs from our sample, at $z$\,\simi4.05, show effective radii up to two times larger than those derived for the HUDF sources at similar redshift. In fact, their sizes are compatible with CEERS galaxies at $z$\,\simi1\,$-$\,3 \citep{Ormerod+24}. Regarding the S\'ersic index, our results are within the range of values shown in the ALMA-selected sample ($n$\,\simi0.4\,$-$\,6) that do not show any evolution with redshift. However, we have not observed any clear trend with the stellar mass.

The fact that two out of our three DSFGs, namely, GN20 and GN20.2a, exhibit larger sizes than field galaxies at similar redshift can be understood as a rapid growth triggered by gravitational interactions arising from their location in an overdense region. Galaxies in proto-clusters have been found to evolve rapidly, with SFR/M$_\mathrm{halo}$ larger than field galaxies \citep{Bai+09,Webb+13,Popesso+15}. Indeed, some hints of a past gravitational interaction have been found in GN20 in both its morphology, with the centroid of the stellar envelope \simi1\,kpc offset from the unresolved nucleus (see Sect.~\ref{subsec:morph-results} and \citealt{Colina+23}) and its H$\alpha$ kinematics, with the presence of non-circular motions and a NW companion at 12\,kpc away and \simi750\,km\,s$^{-1}$ \citep{Ubler+24}. In this context, we offer the interpretation that our three DSFGs are in different gravitational interaction stages. While GN20 and GN20.2a are strong starbursting LTGs that are rapidly growing in mass and size due to a past gravitational interaction, GN20.2b (which resembles a more evolved ETG) is undergoing a `rejuvenisation' process that is likely linked to a more recent accretion of gas that would have enhanced the star formation once again. As GN20.2b presents a more compact structure and a SFR compatible with the SFMS at $z$\,\simi4, it could be understood to be a more evolved galaxy gradually moving from `starburst' to `quiescent' mode before the `rejuvenation' process. This is consistent with the larger fractions of quenching galaxies that have been found at high-mass and high-density environments \citep{Liu+21}. Although BD29079 displays a size ($R_\mathrm{eff}$\,=\,1.52\pmm0.05\,kpc) similar to our most compact DSFG (i.e. GN20.2b), the presence of three bright clumps separated by few kiloparsecs could be interpreted as a sign of an ongoing merger. Similar clumpy structures have been found in simulations and observations of merging systems in their pre-coalescence phases \citep{Ribeiro+17,Calabro+19,Nakazato+24}. Further analysis on the kinematics of these galaxies, tracing possible gravitational interactions, are needed to probe the interpretations given in this section.

When making comparisons with other DSFGs in overdensities discovered at $z$\,$>$3, we can observe that our galaxies present similar physical properties. The SFRs obtained for our sample (i.e.  \simi300\,$-$\,2500\Msyr) are in good agreement with the values derived for the most IR-luminous DSFGs in the overdensity at $z$\,\simi4.57, known as the `distant red core' (i.e. \simi600\,$-$\,2900\Msyr, \citealt{Oteo+18}). Our SFR are also consistent with the values derived for various DSFGs in overdensities at $z$\,\simi2\,$-$\,7 (i.e. \,\simi200\,$-$\,2000\Msyr, \citealt{Walter+12,Pavesi+18,Gomez-Guijarro+19,Alvarez-Marquez+23}). In addition, the typical distances between the members of these overdensities are similar, being usually distributed over an area of a few \simi100\,kpc\,$\times$\,100\,kpc. However, other overdensities at high-$z$ display very different scenarios, where the galaxies are either more concentrated as in A2744-z7p9OD (i.e.  $r$\,\simi6\,kpc, \citealt{Hashimoto+23}) or covering larger areas as in PCl\,J1001+0220 (i.e.  $r$\,\simi2\,Mpc, \citealt{Oteo+18}). 
Another observed scenario of these overdensities forsees only a central object, which is a massive DSFG with SFR\,$>$\,100\,$-$\,200\Msyr, while the rest of the objects are forming stars at a lower rate. This is the case of the $z$\,=\,4.3 overdensity analysed in \citet{Caputi+21}, where the central object is a massive (log(\Mstar/\Ms)\,$>$\,11) star-forming (SFR\,$>$\,220\Msyr) ultra-luminous IR galaxy surrounded by lower mass (7\,$<$\,log(\Mstar/\Ms)\,$<$\,10) starburst galaxies with SFR\,$<$\,20\Msyr. New discoveries of similar overdensities at $z$\,$>$\,3 are still needed to allow us to carry out more exhaustive studies of the physical properties of their galaxies.

\section{Summary}
\label{sec:5.Summary}

This paper presents the first rest-frame near-IR (\simi1.1\,$-$\,3.5\micron) sub-arcsec analysis on the morphology and physical properties for the three DSFGs and the LBG identified in the overdensity around the SMG GN20 at $z$\,\simi4.05. 
Concretely, JWST MIRI imaging in several filters, from F560W to F1800W, have been used to resolve, for the first time, the distribution of their mature stellar population. The high angular resolution (\simi0.25\arcs) of the bluer filters (i.e.  F560W and F770W) allow us to model their spatially resolved stellar brightness distributions and study their non-parametric morphological indices. Finally, a rest-frame far-UV to millimetre SED-fitting analysis is carried out to derive the main physical properties of the sample (e.g.  SFR, \Mstar, sSFR, $A_\mathrm{V}$). The main results of this study are as follows. 

\begin{itemize}

    \item The MIRI images have revealed that our galaxies show a wide range of morphologies in the rest-frame near-IR, from disc+bulge to clump-dominated structures. This near-IR structure contrasts with the rest-frame UV emission for the DSFGs. It appears to be diffuse and, in GN20 and GN20.2a, it is off-centred by \simi4\,kpc with respect to their near-IR counterparts. This offset is understood as a consequence of the extremely heavy obscuration in the nuclear regions of these galaxies. Nevertheless, the LBG galaxy (i.e. \,BD29079) displays a similar structure in the rest-frame UV and near-IR.

    \item We observe that, in the DSFGs, the PAH\,3.3\micron emission traced with the F1800W filter displays more diffuse and less centrally peaked brightness distributions than the rest of the MIRI filters. This indicates that these galaxies are forming stars homogeneously throughout their entire extension, in contrast to what can be derived by looking at the HST images, where the UV emission is only able to leak at the external and less obscured regions.

    \item From the parametric modelling of the stellar brightness distribution observed in MIRI/F560W (rest-frame \simi1.1\micron), we find that GN20 and GN20.2a present disc-like structures recalling spiral galaxies, while GN20.2b shows a very compact light profile with a S\'ersic index $n$\,=\,4.5, consistent with an early-type galaxy. In addition, the three DSFGs present irregular sub-structure in their residual maps. These features, consistent with the structures revealed by the Lucy-Richardson deconvolution, are likely linked to the presence of dust lanes and off-nuclear clumps.

    \item According to the Gini\,$-$\,$M_{20}$ and $C$\,$-$\,$A$ criteria, GN20, GN20.2a and BD29079 are compatible with being late-type galaxies (LTGs), while GN20.2b is an early-type (ETG), consistent with the parametric morphological analysis. The predominance of LTGs is in agreement with the morphological classification of SMGs \citep{Gillman+23} and the general population of galaxies from CEERS \citep{Kartaltepe+23} at z<5.
    
    \item The analysis of the SED  of the galaxies in our sample, reveals that all are massive (\Mstar{}=\,(0.24\,$-$\,1.79)\EXP{11}\Ms), with high SFR (\simi300\,$-$\,2500\Msyr). They also present high internal extinction ($A_\mathrm{V}$\,=\,0.8\,$-$\,1.5\,mag). We observe that the disk-like GN20 is dominating the total star formation of the overdensity with SFR\,=\,2550\pmm150\Msyr, while the compact early-type GN20.2b dominates the stellar mass (\Mstar\,=\,(1.8\pmm0.4)\EXP{11}\Ms). Finally, we find that although BD29079 (LBG) is as massive as GN20.2a, it is forming stars at the lowest rate among the sample (310\pmm80\Ms\,yr$^{-1}$). 
    
    \item The galaxies classified as LTG (GN20, GN20.2a and BD29079) are all above the main sequence of star-forming galaxies at z\,\simi4 by $>$\,0.5\,dex, while GN20.2b, classified as ETG, is compatible with the high-mass end of the main sequence. When comparing with larger samples of galaxies at this redshift, our galaxies are very similar to the ALMA-selected dusty optically-faint galaxies from \citet{Xiao+23} and well above the [CII] emitters from the ALPINE sample on the SFR$-$\Mstar plane.
    
    \item The $R_\mathrm{eff}$ values derived for our DSFGs are up to two times larger (i.e.  \simi1.3\,$-$\,3.5\,kpc) than expected at their redshift, according to recent JWST studies based on CEERS and ALMA-HUDF galaxies. These large values can be understood as result of a rapid growth in mass and size as a consequence of being located in an overdensity. In this scenario, GN20.2b (with the highest stellar mass, lowest gas mass and lowest sSFR) seems to be in a more advanced stage of this evolution than GN20 and GN20.2a.

    \item When making comparisons with other DSFGs in overdensities at $z$\,\simi2\,$-$\,7, our targets display similar SFRs (\simi300\,$-$\,2500\Msyr), depletion times ($<$\,100\,Myr) and projected separations (\simi1\,$-$\,200\,kpc) values. These results support a scenario where galaxy overdensities play a key role in triggering the extreme DSFGs in the early Universe. 

\end{itemize}

In this work, we have shown how the JWST/MIRI images, tracing the rest-frame near-IR, have unveiled the true morphology of the stellar component for the three DSFGs and the LBG present in the overdensity at $z$\,=\,4.05. These MIRI images have revealed that their extended and (in two out of the three DSFGs) disc-like morphologies stand in contrast to the diffuse and offset rest-frame UV emission that had misled our understanding of their stellar distribution. However, the origin of the differences in their morphologies in the near-IR is still uncertain. Further observations of their optical counterparts are therefore needed to fill the gap between the UV and near-IR emission and help improve our understanding of their intrinsic structure and the physical processes occurring in these objects.

\begin{acknowledgements}
We thank the anonymous referee for helpful and constructive comments which improved this paper. 
A.C.G., J.A-M., L.C. and A.L. acknowledge support by grant PIB2021-127718NB-100, P.G.P-G. and L.C. by grant PGC2018-093499-B-I00. 
A.A-H. acknowledges support from grant PID2021-124665NB-I00 from the Spanish Ministry of Science and Innovation/State Agency of Research MCIN/AEI/10.13039/501100011033 and by “ERDF A way of making Europe”. 
L.C. acknowledges support from the fellowship from the ``la Caixa” Foundation (ID 100010434) with code LCF/BQ/PR24/12050015. L.B. and F.W. acknowledge support from the ERC Advanced Grant 740246 (Cosmic\_Gas).
A.B., J.M \& G.\"O. acknowledge support from the Swedish National Space Administration (SNSA).
O.I. acknowledges the funding of the French Agence Nationale de la Recherche for the project iMAGE (grant ANR-22-CE31-0007),
J.H. and D.L. were supported by a VILLUM FONDEN Investigator grant (project number 16599). This work was also supported by a research grant (VIL54489) from VILLUM FONDEN.
K.I.C. acknowledges funding from the Netherlands Research School for Astronomy (NOVA) and from the Dutch Research Council (NWO) through the award of the Vici Grant VI.C.212.036.
J.P.P. and T.T. acknowledge financial support from the UK Science and Technology Facilities Council, and the UK Space Agency.
T.P.R. would like to acknowledge support from the ERC under advanced grant 743029 (EASY). The Cosmic Dawn Center (DAWN) is funded by the Danish National Research Foundation under grant No. 140.
SG acknowledges financial support from the Villum Young Investigator grant 37440 and 13160 and the Cosmic Dawn Center (DAWN), funded by the Danish National Research Foundation under grant DNRF140.
L.B. and F.W. acknowledge support by ERC AdG grant 740246 (Cosmic-Gas).

The authors thank Jacqueline Hodge for providing the CO data of GN20.
The work presented is the effort of the entire MIRI team and the enthusiasm within the MIRI partnership is a significant factor in its success. MIRI draws on the scientific and technical expertise of the following organisations: Ames Research Center, USA; Airbus Defence and Space, UK; CEA-Irfu, Saclay, France; Centre Spatial de Liége, Belgium; Consejo Superior de Investigaciones Científicas, Spain; Carl Zeiss Optronics, Germany; Chalmers University of Technology, Sweden; Danish Space Research Institute, Denmark; Dublin Institute for Advanced Studies, Ireland; European Space Agency, Netherlands; ETCA, Belgium; ETH Zurich, Switzerland; Goddard Space Flight Center, USA; Institute d'Astrophysique Spatiale, France; Instituto Nacional de Técnica Aeroespacial, Spain; Institute for Astronomy, Edinburgh, UK; Jet Propulsion Laboratory, USA; Laboratoire d'Astrophysique de Marseille (LAM), France; Leiden University, Netherlands; Lockheed Advanced Technology Center (USA); NOVA Opt-IR group at Dwingeloo, Netherlands; Northrop Grumman, USA; Max-Planck Institut für Astronomie (MPIA), Heidelberg, Germany; Laboratoire d’Etudes Spatiales et d'Instrumentation en Astrophysique (LESIA), France; Paul Scherrer Institut, Switzerland; Raytheon Vision Systems, USA; RUAG Aerospace, Switzerland; Rutherford Appleton Laboratory (RAL Space), UK; Space Telescope Science Institute, USA; Toegepast- Natuurwetenschappelijk Onderzoek (TNO-TPD), Netherlands; UK Astronomy Technology Centre, UK; University College London, UK; University of Amsterdam, Netherlands; University of Arizona, USA; University of Cardiff, UK; University of Cologne, Germany; University of Ghent; University of Groningen, Netherlands; University of Leicester, UK; University of Leuven, Belgium; University of Stockholm, Sweden; Utah State University, USA. A portion of this work was carried out at the Jet Propulsion Laboratory, California Institute of Technology, under a contract with the National Aeronautics and Space Administration. We would like to thank the following National and International Funding Agencies for their support of the MIRI development: NASA; ESA; Belgian Science Policy Office; Centre Nationale D'Etudes Spatiales (CNES); Danish National Space Centre; Deutsches Zentrum fur Luft-und Raumfahrt (DLR); Enterprise Ireland; Ministerio De Econom\'ia y Competitividad; Netherlands Research School for Astronomy (NOVA); Netherlands Organisation for Scientific Research (NWO); Science and Technology Facilities Council; Swiss Space Office; Swedish National Space Board; UK Space Agency. 

This work is based on observations made with the NASA/ESA/CSA James Webb Space Telescope. The data were obtained from the Mikulski Archive for Space Telescopes at the Space Telescope Science Institute, which is operated by the Association of Universities for Research in Astronomy, Inc., under NASA contract NAS 5-03127 for \textit{JWST}; and from the \href{https://jwst.esac.esa.int/archive/}{European \textit{JWST} archive (e\textit{JWST})} operated by the ESDC.

\end{acknowledgements}

\bibliographystyle{aa} 
\bibliography{bibliography.bib} 

\appendix

\section{Photometry used in the SED fitting}
\label{app:Photometry}

Table~\ref{tab:fluxes} displays the aperture photometry obtained for HST and JWST as explained in Sect.~\ref{subsec:sed_fitting}, along with the ancillary far-IR and sub-mm data used during the SED fitting.

\begin{table*}[!ht]
    \centering
\caption{Photometry used in the SED fitting analysis}
\begin{tabular}{cccccc}
\hline
          Filter & units &                  GN20 &                GN20.2a &                GN20.2b &              BD29079 \\
\hline
 $z $&  &              4.0553\pmm0.0002$^\dagger$  &       4.0508\pmm0.0013$^\dagger$  &     4.0563\pmm0.0003$^\dagger$  &            4.058$^{\dagger\dagger}$ \\
\hline 
ACS/F435W &  $\mu$Jy &               $<$\,0.092 &               $<$\,0.086 &               $<$\,0.081 &               $<$\,0.061 \\
       ACS/F606W &  $\mu$Jy &   0.348\,$\pm$\,0.027 &   0.290\,$\pm$\,0.025 &   0.152\,$\pm$\,0.023 &   0.306\,$\pm$\,0.017 \\
       ACS/F775W &  $\mu$Jy &   0.710\,$\pm$\,0.041 &   0.657\,$\pm$\,0.039 &   0.308\,$\pm$\,0.036 &   0.875\,$\pm$\,0.027 \\
       ACS/F814W &  $\mu$Jy &   0.758\,$\pm$\,0.050 &   0.573\,$\pm$\,0.047 &   0.225\,$\pm$\,0.043 &   1.047\,$\pm$\,0.035 \\
      ACS/F850LP &  $\mu$Jy &   0.822\,$\pm$\,0.038 &   0.628\,$\pm$\,0.037 &   0.281\,$\pm$\,0.034 &   1.159\,$\pm$\,0.025 \\
      WFC3/F105W &  $\mu$Jy &   0.905\,$\pm$\,0.062 &   0.608\,$\pm$\,0.060 &   0.360\,$\pm$\,0.054 &   1.202\,$\pm$\,0.040 \\
      WFC3/F125W &  $\mu$Jy &   1.060\,$\pm$\,0.083 &   0.531\,$\pm$\,0.082 &   0.456\,$\pm$\,0.073 &   1.373\,$\pm$\,0.040 \\
      WFC3/F140W &  $\mu$Jy &   1.21\,$\pm$\,0.11 &                     - &   0.455\,$\pm$\,0.094 &   1.333\,$\pm$\,0.067 \\
      WFC3/F160W &  $\mu$Jy &   1.148\,$\pm$\,0.083 &   0.760\,$\pm$\,0.076 &   0.603\,$\pm$\,0.068 &   1.470\,$\pm$\,0.039 \\
     MIRIM/F560W &  $\mu$Jy &  10.76\,$\pm$\,0.23 &   3.32\,$\pm$\,0.23 &   8.17\,$\pm$\,0.25 &   2.47\,$\pm$\,0.15 \\
     MIRIM/F770W &  $\mu$Jy &  17.85\,$\pm$\,0.30 &   6.07\,$\pm$\,0.28 &  12.54\,$\pm$\,0.32 &   2.86\,$\pm$\,0.18 \\
    MIRIM/F1280W &  $\mu$Jy &  23.48\,$\pm$\,0.73 &   9.01\,$\pm$\,0.72 &  10.97\,$\pm$\,0.79 &   6.97\,$\pm$\,0.47 \\
    MIRIM/F1800W &  $\mu$Jy &  35.8\,$\pm$\,3.6 &  13.9\,$\pm$\,3.4 &  14.3\,$\pm$\,3.7 &  16.2\,$\pm$\,2.4 \\
  IRAC/3.6$\mu$m &  $\mu$Jy &   5.949\,$\pm$\,0.090 &   2.48\,$\pm$\,0.10 &   4.50\,$\pm$\,0.12 &   4.29\,$\pm$\,0.04 \\
  IRAC/4.5$\mu$m &  $\mu$Jy &   8.249\,$\pm$\,0.077 &   2.637\,$\pm$\,0.085 &   6.42\,$\pm$\,0.09 &   3.08\,$\pm$\,0.07 \\
  IRAC/5.7$\mu$m &  $\mu$Jy &  14.29\,$\pm$\,0.69 &   4.09\,$\pm$\,0.76 &   9.03\,$\pm$\,0.84 &   3.35\,$\pm$\,0.50 \\
  IRAC/7.8$\mu$m &  $\mu$Jy &  22.57\,$\pm$\,0.55 &   6.80\,$\pm$\,0.60 &  13.37\,$\pm$\,0.70 &   4.69\,$\pm$\,0.52 \\
   MIPS/24$\mu$m &  $\mu$Jy &  65.5\,$\pm$\,3.5 &  30.2\,$\pm$\,5.6 &  12.0\,$\pm$\,4.3 &  47.3\,$\pm$\,5.4 \\
  PACS/100$\mu$m &      mJy &   0.70\,$\pm$\,0.42 &   0.12\,$\pm$\,0.44 &   0.61\,$\pm$\,0.46 &   0.54\,$\pm$\,0.33\\
  PACS/160$\mu$m &      mJy &   5.5\,$\pm$\,1.0 &   1.2\,$\pm$\,1.4 &   2.3\,$\pm$\,1.4 &   1.72\,$\pm$\,0.65 \\
 SPIRE/250$\mu$m &      mJy &  18.6\,$\pm$\,2.7 &                     - &                     - &               $<$\,5.4 \\
 SPIRE/350$\mu$m &      mJy &  41.3\,$\pm$\,5.2 &                     - &                     - &                     - \\
 SPIRE/500$\mu$m &      mJy &  39.7\,$\pm$\,6.1 &                     - &                     - &               $<$\,6.6 \\
 SCUBA/850$\mu$m &      mJy &  20.3\,$\pm$\,2.1 &   9.9\,$\pm$\,2.3 &   9.9\,$\pm$\,2.0 &                     - \\
      PdBI/1.2mm &      mJy &   8.47\,$\pm$\,0.79 &   3.83\,$\pm$\,0.48 &   3.25\,$\pm$\,0.52 &                     - \\
      PdBI/2.2mm &      mJy &   0.95\,$\pm$\,0.14 &   0.52\,$\pm$\,0.23 &               $<$\,0.80 &                     - \\
      PdBI/3.3mm &      mJy &   0.229\,$\pm$\,0.036 &   0.177\,$\pm$\,0.071 &   0.114\,$\pm$\,0.035 &                     - \\
\end{tabular}
    \label{tab:fluxes}
    \tablefoot{HST and MIRI fluxes measured through aperture photometry (see Sect.~\ref{subsec:sed_fitting}). Photometric values beyond 2\micron correspond to the GOODS-N catalogue presented in \citet{Liu+18}. Upper limits are set as 3$\sigma$ values in those filters where the uncertainty is larger than the flux measurement. Hyphens represent the non-coverage of the object by the corresponding filter. \,$^\dagger$ Redshift assumed during the SED fitting, computed in \citet{Tan+14}. \,$^{\dagger\dagger}$ Redshift derived in \citet{Daddi+09}. }
\end{table*}

\section{Robustness of the applied high-pass filter}
\label{ref:high_pass_appendix}


As mentioned in Sect. \ref{subsec:morph-results} (and made all the more evident in Fig. \ref{fig:lenstronomy}), our model does not recover all the substructures of the sample. For example, the northern emission of the central region of BD29079 reveals two blobs east and west, which are visible in the residuals and the high-pass filtered images. Critically, we want to explore the validity of the Lucy-Richardson algorithm used by introducing additional noise to the data. Adjusting the intensity of the noise, we will probe the robustness of the detected emission, which is only visible in the residuals and the high-pass filtered images (Fig. \ref{fig:lenstronomy}). 
To provide a better understanding of the general impact of the noise, we measured the median emission of the background at a distance of about 10 kpc away from the central emission of every galaxy in the sample and compared the variance $P_i(k)$\,=\,$(i^k/k!)$\,$e^{-i}$ of the input data and the filtered results. We normalised the data to unity and list the noise before and after high-pass filtering in Table \ref{tab:noise}.

\begin{table*}[]
    \centering
    \caption{Properties of the data}
\begin{tabular}{ccccc}
\hline \hline
  Galaxy & Noise - Normalised data & Data Noise Variance & Noise - Filtered results & Filtered Noise Variance \\
         & $10^{-3}\,\%$  &  & $10^{-3}\,\%$  &  \\
  \hline
 GN20    &     23.31\pmm 8.70 & 0.0012 &  11.98\pmm8.67  & 0.0012 \\
 GN20.2a &     32.39\pmm30.44 & 0.0015 &   8.13\pmm5.92  & 0.0016 \\
 GN20.2b &     40.51\pmm40.26 & 0.0005 &   2.92\pmm2.61  & 0.0009 \\
 BD29079 &     25.01\pmm16.19 & 0.0007 &   5.74\pmm4.93  & 0.0002 \\
 \hline
 \end{tabular}
    \label{tab:noise}
  \tablefoot{Noise measurements of the input data and the high-pass filtered images. The indicated uncertainties are related to the standard deviation. Fluctuations of the noise are expected due to the nature of the background emission.}  
\end{table*}

\begin{figure*}[!ht]
\centering
   \includegraphics[width=\textwidth]{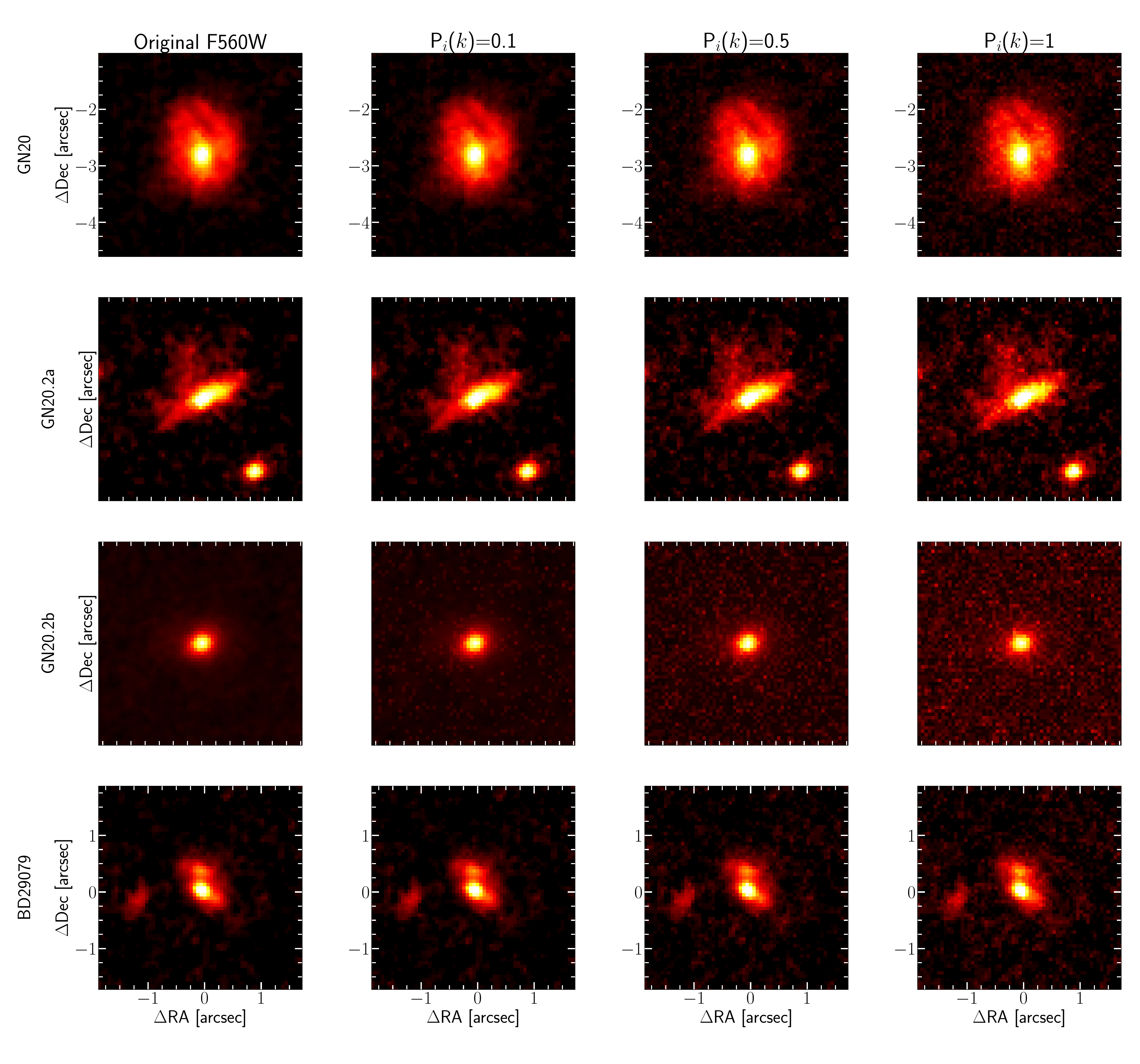}
      \caption{Comparison of the original data (left column) and the Poisson noise added versions that are classified by the related variance P$_i$($k$). Lower values of P$_i$($k$) are associated with less noise. The variance of the input data is listed in Table \ref{tab:noise}.}
         \label{fig:input_data_comp}
\end{figure*}

We use the listed values as an input for noise model that we add to the data. For consistency, we replicate every step as described in Sect. \ref{sec:4.Results} to reproduce the high-pass filtered results. To test the impact of noise, we use a Poission model that represents different background variations.
For this, we measure the variance of the noise, which serves as the intensity quantity. This quantity is used for the Poisson noise that is added to the data. Due to the low measured noise level and the related estimated variance, we manually set the variance to three different values (0.1, 0.5, 1.0), which exceeds the measured noise variance by several magnitudes. In Fig. \ref{fig:input_data_comp}, we display the observations and the added Poisson noise data which is then filtered with the LR algorithm.

\begin{figure*}[!ht]
\centering
   \includegraphics[width=\textwidth]{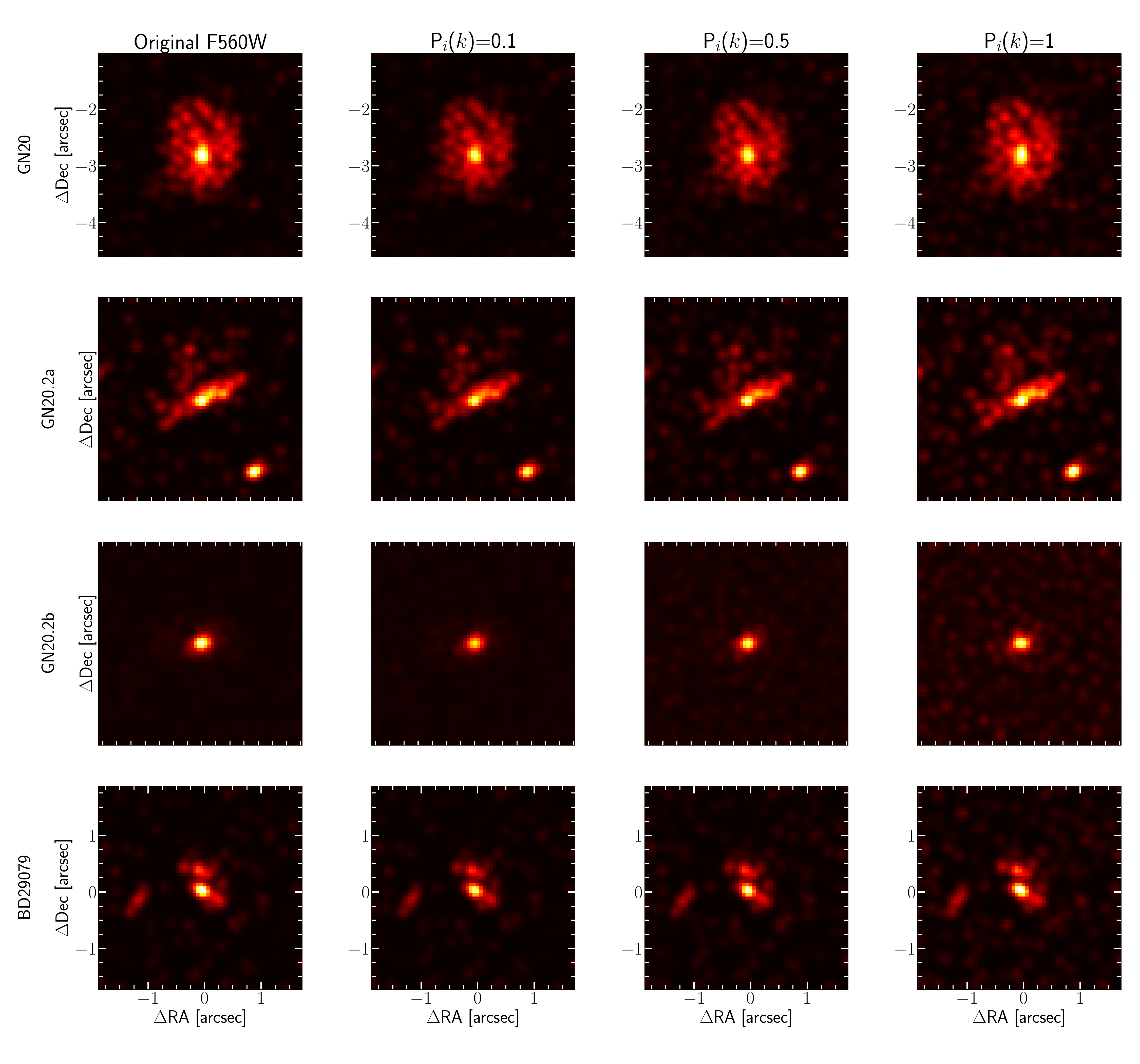}
      \caption{Lucy-Richardson high-pass filtered results for the original and mocked data shown in Fig. \ref{fig:input_data_comp}.}
         \label{fig:filtered_noise_data_comp}
\end{figure*}

As shown in Fig. \ref{fig:filtered_noise_data_comp}, we find no significant emission above the noise level, although we used a variance several magnitudes above the measured values (Table \ref{tab:noise}). From this analysis it is evident that no significant structures are produced by the LR algorithm.

This is expected because the spatial frequency of the noise is below the dimension of the PSF. In other words, the algorithm does not produce an observable emission from sporadic noise fluctuations \citep{Peissker2020}.


\end{document}